\documentclass[aip,pof]{revtex4-2}

\usepackage{color}
\usepackage{graphicx}
\providecommand\be{\begin{equation}}
\providecommand\ee{\end{equation}}
\providecommand\bsub{\begin{subequations}}
\providecommand\esub{\end{subequations}}

\usepackage{hyperref}

\begin{document}


\title{Torricelli's Curtain: Morphology of Horizontal Laminar Jets Under Gravity} 



\author{O. Tramis}
\author{E. Merlin-Anglade}
\author{G. Paternoster}
\author{M. Rabaud}
\email[]{marc.rabaud@universite-paris-saclay.fr}
\author{N. M. Ribe}
\email[]{neil.ribe@universite-paris-saclay.fr}
\affiliation{Lab FAST. Universit\'e Paris-Saclay, CNRS, 91405 Orsay, France}


\date{\today}

\begin{abstract}
Viscous fluid exiting a long horizontal circular pipe develops a complex structure 
comprising a primary jet above and a smaller secondary jet below with a thin
fluid curtain connecting them. We present here a combined experimental,
theoretical and numerical study of this `Torricelli's curtain' phenomenon, focusing on
the factors that control its morphology.  The dimensional  parameters
that define the problem are the pipe radius $a$, the mean exit velocity
$U$ of the fluid, the gravitational acceleration $g$,
and the fluid's density $\rho$, kinematic viscosity $\nu$ and coefficient 
of surface tension $\gamma$. Rescaling of experimentally
measured trajectories of the primary and secondary jets using
$a$ for the vertical coordinate and $L_D = U (a/g)^{1/2}$ for
the horizontal coordinate $x$ collapses the data onto universal
curves for $x < 10 L_D$. We propose a theoretical model for the curtain
in which particle trajectories result from the composition of two
motions: a horizontal component corresponding to the evolving
axial velocity profile of an axisymmetric viscous jet, and a
vertical component due to free fall under gravity. The model predicts well the trajectory of 
the primary jet, but somewhat less well that of the secondary jet.
We suggest that the remaining discrepancy may be explained by
surface tension-driven (Taylor-Culick) retraction of the secondary
jet. Finally, direct numerical simulation 
reveals recirculating `Dean' vortices in vertical sections of the primary
jet, placing Torricelli's curtain firmly within the context of flow in 
curved pipes. 

\end{abstract}


\maketitle 


\section{Introduction}
\label{introduction}

\begin{figure*}
\includegraphics[width=0.8\linewidth]{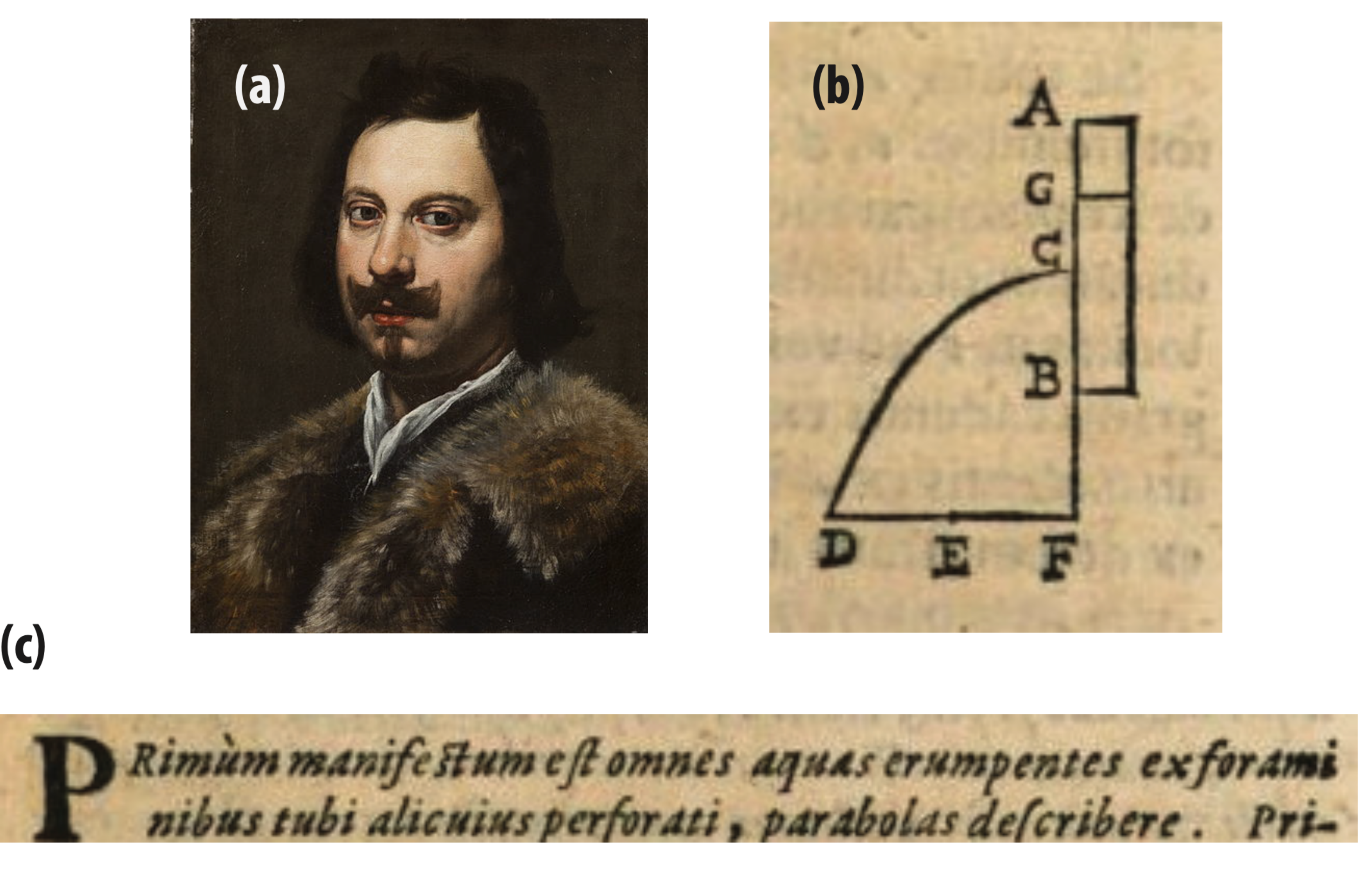}
\caption{\label{fig_torricelli}
Torricelli's (1644) work on the shape of horizontal jets.  (a)
Portrait of Torricelli by Lorenzo Lippi (ca. 1647). (b) Torricelli's
sketch of a vertical tube AB filled with water to the level G and perforated
at C. CD is the parabolic trajectory of the jet. (c) Torricelli's 
enunciation of the parabolic trajectories of water jets. Source for
part (a): ``Evangelista Torricelli", Wikipedia, Wikimedia Foundation, 20 January 2021, \url{https://en.wikipedia.org/wiki/Evangelista_Torricelli}.
Source for parts (b) and (c): \url{https://books.google.fr/books?id=UQsOAAAAQAAJ}.}
\end{figure*}

One of the oldest problems in fluid mechanics is to determine the 
form taken by a jet of water issuing horizontally from a hole pierced
in the side of a water-filled container. The solution was first discovered
by Evangelista Torricelli \citep{torricelli1644} (fig. \ref{fig_torricelli}a) . 
Torricelli's work on horizontal water jets is found in Book 2 of his treatise
``On the Motion of Naturally Descending Heavy Bodies", in a section 
entitled ``On the Motion of Waters".  
He states his conclusion as follows (fig. \ref{fig_torricelli}c): ``In the first place it is evident
that all waters issuing from holes in any perforated tube describe parabolas". 
Torricelli's discovery of the parabolic trajectories of water jets is one of the 
founding results of the science of fluid mechanics.
Dorfman \cite{dorfman58} provides an accessible introduction to Torricelli's life and work.

\begin{figure}
\includegraphics[width=0.8\linewidth]{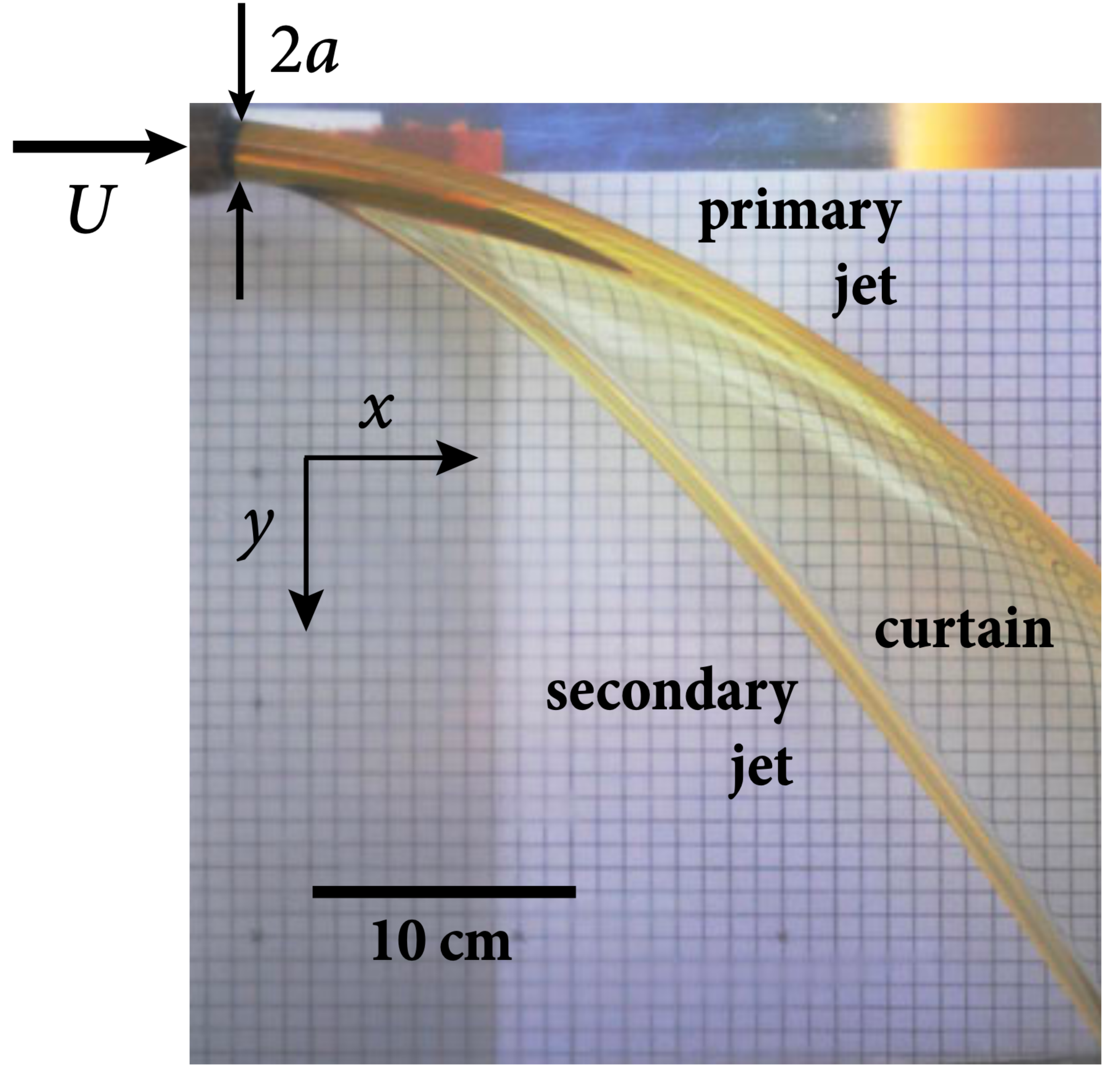}
\caption{\label{fig_structure}
Structure of Torricelli's curtain. A jet of mineral oil is ejected horizontally
at mean speed $U = 0.97$ m s$^{-1}$ from a 6 m long pipe with inner radius $a = 9.5$ mm. The
initial jet splits into a primary jet above and a smaller secondary jet below with 
a thin fluid curtain between them. For typical horizontal and vertical cross-sections 
of the whole structure see fig. \ref{fig_silicone37_gerris_hxy}.
}
\end{figure}

Why revisit now the nearly 400 year-old question of the shape of horizontal jets?
Our renewed interest in this problem began with a surprising observation
in a teaching laboratory setup designed to illustrate the properties of 
Poiseuille flow. The working fluid is mineral oil, which is pumped
through a long (6 m) pipe of inner radius 9.5 mm equipped with pressure sensors at fixed intervals.
The flowing oil exits the downstream end of the pipe into a transparent air-filled 
chamber, whence it is recycled back to the upstream end. In view of 
Torricelli's well-known result, we expected the jet exiting the pipe to have a
parabolic trajectory. What we saw instead was the complex
structure shown in fig. \ref{fig_structure}. The initially round jet rapidly splits
into a primary jet above and a smaller secondary jet below with a thin
fluid curtain connecting them. We baptized this phenomenon 
``Torricelli's curtain" in honor of Torricelli's seminal contributions to our
understanding of horizontal liquid jets. 

\begin{figure}
\includegraphics[width=0.8\linewidth]{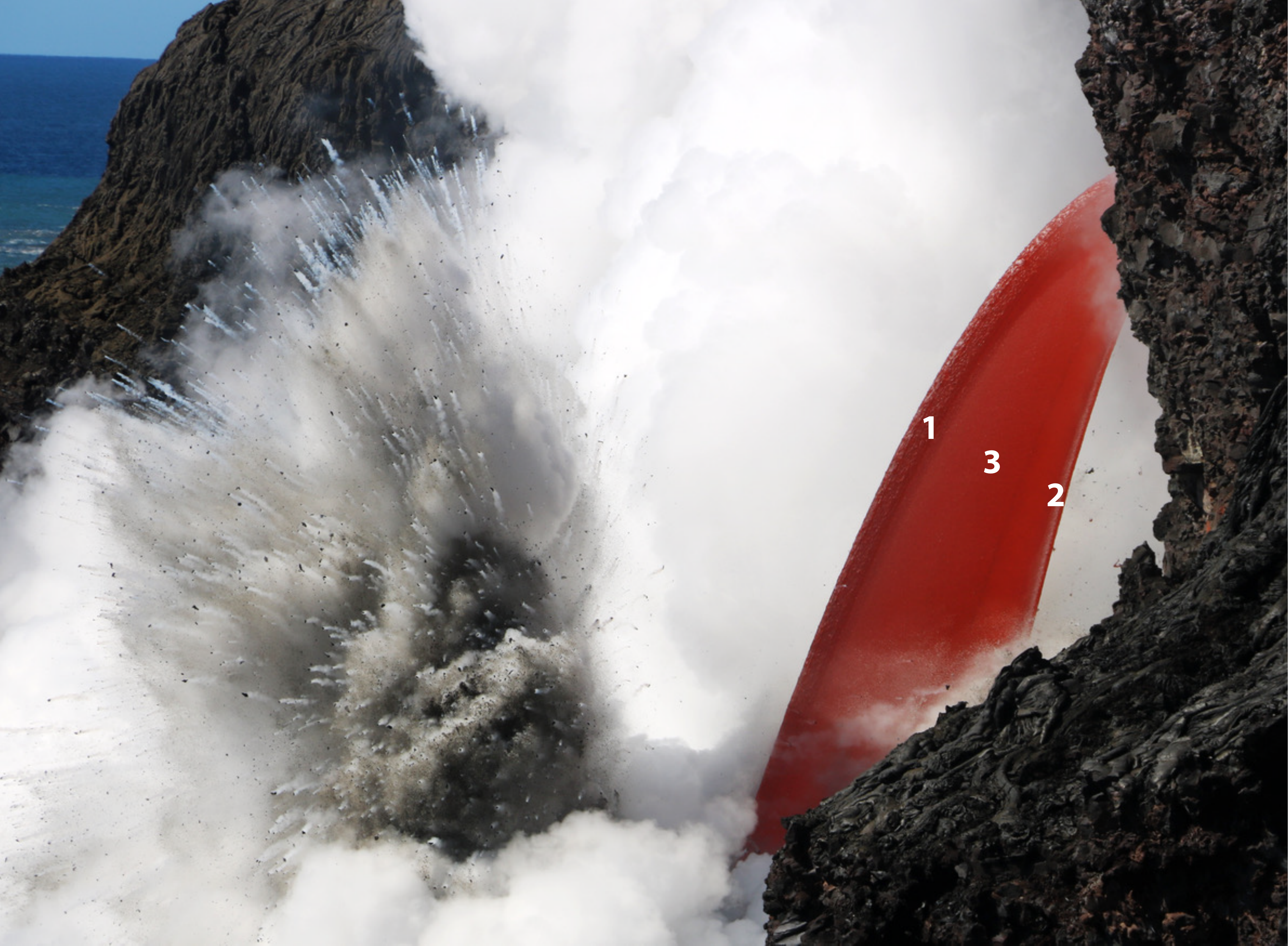}
\caption{\label{fig_hawaii}
A natural example of Torricelli's curtain, formed by a horizontal jet of molten lava
on the flank of Kilauea volcano, Hawaii on 28 January 2017. The distance from the
outlet to the sea surface is about 28 m. The numbers 1, 2 and 3 indicate the 
primary jet, the secondary jet, and the curtain, respectively. Public
domain source:
\url{https://www.usgs.gov/media/images/open-lava-stream-continues-ocean-entry}.
Credit: U.S. Geological Survey, Department of the Interior/USGS.}
\end{figure}

We soon discovered that Torricelli's curtain was not confined to our laboratory.
The phenomenon can be seen in the pedagogical movie``Turbulence" of the National 
Committee for Fluid Mechanics Films  (\url{http://web.mit.edu/hml/ncfmf.html}) 
when an increase in the liquid viscosity causes the flow to 
become laminar.
It can also be found in Nature. Fig. \ref{fig_hawaii} shows an example
of what volcanologists call a `firehose':  an initially horizontal
jet of molten lava falling into the ocean. The primary jet is
visible as a band of darker red just below the upper extremity of the 
curtain. 

With its two jets connected by a thin sheet, Torricelli's curtain obviously involves 
separate pieces with different dynamics. The literature on both jets and sheets
is immense, and a complete survey would require a separate paper. Instead, to 
focus the discussion we concentrate on the aspects
most relevant to our work: steady axisymmetric laminar capillary jets exiting long pipes at high
Reynolds number $Re$, and planar fluid sheets. Finally, we briefly survey past
experimental work on buoyant jets injected horizontally into a quiescent fluid with
a similar viscosity, a system that displays some striking similarities to Torricelli's curtain. 

All beginning students of fluid mechanics know that steady laminar flow in a long pipe
has a parabolic (Poiseuille) velocity profile with zero velocity but finite shear stress at the pipe's inner wall.
The maximum velocity at the center of the pipe is twice the mean velocity $U$.
When the liquid exits the pipe into air, the boundary condition on its outer surface changes abruptly
from one of zero velocity to one of zero shear stress. In the absence of gravity and surface
tension, the resulting mismatch
of the shear stress between the surface and the interior is gradually erased by 
radial viscous diffusion, resulting in the establishment of a uniform (plug flow) profile
across the jet. The far-field velocity $U_{\infty}= (4/3)U$ and radius $a_{\infty} =  (\sqrt{3}/2)a$ of the jet
are determined by the requirements of conservation of volume flux and momentum flux.
This plug flow is established at a downstream distance $x \approx 0.25 a Re$, where
$Re= Ua/\nu$ is the Reynolds number.

At distances $x\ll 0.25 a Re$, radial viscous diffusion is confined
to a thin boundary layer of thickness $\delta(x)\ll a$ adjoining the surface of the jet,
and has not yet been influenced by the jet's finite radius. The structure of 
the boundary layer can therefore be treated as a two-dimensional problem
of parabolic type with an outer boundary condition of constant shear $\dot\gamma$
given by the near-wall Poiseuille velocity profile. This problem was solved
by Goren \cite{goren66jfma}, who found a self-similar solution in which  
$\delta \sim (\nu x/\dot\gamma)^{1/3}$.
Goren \cite{goren66jfmb} determined the shape of an axisymmetric jet far downstream by linearizing
the Navier-Stokes equations about the jet's final uniform state at $x/(a Re)\to\infty$.
Subsequent studies considered vertical jets subject to axial gravity and
(in most cases) surface tension. Examples include
Duda and Vrentas\cite{duda67ces} and Tsukiji and Takahashi\cite{tsukiji87jsme}, who transformed the boundary-layer form
of the Navier-Stokes equations
to streamline coordinates and solved them numerically
to obtain the structure of the jet at all distances $x/(a Re)$.
Philippe and Dumargue\cite{philippe91zamp} used matched asymptotic expansions to obtain a
semi-analytical solution for the jet structure, but did not consider the effect of surface
tension. 
Oguz\cite{oguz98pof} solved the boundary-layer equations for an axisymmetric jet
without surface tension using eigenfunction expansions and a Galerkin numerical method.
Finally, in the context of a stability analysis Sevilla\cite{sevilla11} determined the
steady state of a high-$Re$ capillary jet using a numerical method of lines
due to Gordillo \textit{et al.}\cite{gordillo01}, which we too shall use in the sequel. 

Turning to vertically flowing fluid sheets, we first note a seminal contribution of
G. I. Taylor, who in the Appendix to Brown\cite{brown61jfm} derived a one-dimensional
equation for the steady-state velocity of a falling liquid sheet governed by a balance of 
gravity, inertia and extensional viscous forces. Taylor derived his equation assuming
that the velocity was constant across the sheet (plug flow). The equation was later
generalized to include the effects of surface tension and nonstationarity
\citep{erneux93pof, brenner99pof}. For our purposes, the most interesting 
phenomena connected with liquid sheets are those that occur near a free edge
retracting under the influence of surface tension.  
Taylor\cite{taylor59procrsoc} and Culick\cite{culick60jap} considered the retraction
of the edge of an effectively inviscid liquid sheet with thickness $h$ in the absence
of gravity. By balancing
surface tension and inertia, they found that the velocity of retraction is $U_{tc} = (2\gamma/\rho h)^{1/2}$
where $\gamma$ is the coefficient of surface tension and $\rho$ is the density.
Keller\cite{keller83pof} and Keller \textit{et al.}\cite{keller95pof} 
analyzed the shape of the growing rim on
the edge of a retracting inviscid sheet and found it to be a cylinder whose radius
increases as the square root of the time. 
Brenner and Gueyffier\cite{brenner99pof} solved numerically the one-dimensional thin-sheet equations 
mentioned previously. They found three different regimes  depending on the 
Ohnesorge number $Oh = \nu (\gamma h/\rho)^{-1/2}$ and the 
Reynolds number $Re_f = V L/\nu$, where $L$ is the lateral extent of 
the sheet perpendicular to its edge. Both regimes with $Re_f\gg 1$ show the formation of a growing rim,
with inward-propagating capillary waves when $Oh\ll 1$.  
Song and Tryggvason\cite{song99pof} and 
S\"underhauf \textit{et al.}\cite{suenderhauf02pof}  
solved the full two-dimensional Navier-Stokes equations
in a thin sheet geometry to investigate rim formation. Both studies found that for
$Oh\ll 1$ there is a growing quasi-cylindrical rim with inward-propagating capillary waves.
Roche \textit{et al.}\cite{roche06pof} studied experimentally the shape of the hole 
downstream of a needle piercing a vertically flowing liquid curtain, and found good agreement
with a theoretical model including the effect of surface tension acting on the rim. 
Savva and Bush\cite{savva07jfm} performed a theoretical and numerical study of viscous sheet retraction
and deduced new analytical expressions for the retraction speed at rupture and the 
evolution of the maximum sheet thickness.
Gordillo \textit{et al.}\cite{gordillo11pof} presented both two-dimensional and
one-dimensional numerical solutions for retracting sheets, and determined an analytical
solution of the one-dimensional equations that is valid in the asymptotic limit of large times. 
This is also an appropriate place to mention recent theoretical work by Benilov \citep{benilov21jfm} on
the related problem of a two-dimensional liquid curtain with strong surface tension
ejected from a horizontal slot in a field of gravity.

Finally, there have been a number of experimental studies of
buoyant jets injected horizontally into a quiescent ambient fluid, the source
of the buoyancy being either a temperature difference \citep{anwar72lhb, satyanarayana82ijhmt}
or a compositional one \citep{arakeri00jfm, querzoli05jhr, deri11pof, shao17ijhmt}.
These experiments are similar to the configuration of Torricelli's curtain
in that gravity acts normal to the jet axis, but different in that the jet and the
ambient fluid have comparable viscosities and no surface tension. In the 
context of our work the observations of Arakeri \textit{et al.}\cite{arakeri00jfm} are 
particularly noteworthy. These authors performed their experiments
by injecting pure water jets into denser brine solutions, so that the effective
gravitational  force is directed upwards. They observed that in many cases 
the jet bifurcated into a primary jet and a  rising ``plume" in the form
of a thin sheet parallel to the flow direction.  In fact, Fig. 8 of 
Arakeri \textit{et al.}\cite{arakeri00jfm}
shows a shadowgraph image that 
looks very much like an upside-down version of Torricelli's curtain, with
primary and secondary jets connected by a (presumably) much thinner vertical sheet
of the same fluid.  The observations
of Arakeri \textit{et al.}\cite{arakeri00jfm} were broadly confirmed by subsequent work
\cite{querzoli05jhr, deri11pof,shao17ijhmt}. 

\section{Dimensional analysis}
\label{sec_dimanal}

As a prelude to our subsequent investigations, we
use dimensional analysis to determine the dimensionless
groups governing Torricelli's curtain. The parameters of the experiment
are the fluid density $\rho$, the kinematic viscosity $\nu$, the 
coefficient of surface tension $\gamma$, the pipe radius $a$, the 
mean exit velocity $U$, and the gravitational acceleration
$g$. Of these six parameters, three have independent dimensions.
Buckingham's $\Pi$-theorem then tells us that three independent
dimensionless groups can be formed from the six dimensional
parameters. To help us choose the definitions of
these groups, we note two facts. First, a typical experiment
consists in varying $U$ for fixed values of the other dimensional 
parameters. This suggests that $U$ ought to appear in only
one dimensionless group, which we choose to be the Reynolds number
\be
Re = \frac{U a}{\nu}
\ee
which measures the ratio of inertia to viscous forces. The second 
fact is that the curtain can be considered to be due to gravity
acting on a jet that would otherwise have been horizontal and axisymmetric. 
This suggests that $g$ should appear in only one dimensionless
group. We choose this group to be the ratio of two characteristic 
length scales. The first is the `Dean length' $L_D$, the distance from
the pipe exit at which a fluid particle moving with a horizontal speed $\sim U$
has fallen a distance $\sim a$ under gravity. Assuming a ballistic
(parabolic) trajectory, we find 
\be
L_D = U (a/g)^{1/2}.
\label{deanlength}
\ee
The
second lengthscale $L_{\nu} = a Re$ is the distance from the pipe exit at which
radial viscous diffusion over a length $\sim a$ has occurred. Our 
second dimensionless group is thus the `Dean number' $De = L_{\nu}/L_D$, or
\be
De = \left(\frac{a^3 g}{\nu^2}\right)^{1/2}.
\label{pig}
\ee
The third dimensionless group measures the effect of surface
tension, and should contain neither $U$ nor $g$. This group is the
Laplace number 
\be
La = \frac{\gamma a}{\rho \nu^2} \equiv \frac{(\gamma/a) (\rho U^2)}{(\rho\nu U/a)^2}.
\label{laplace}
\ee
As (\ref{laplace}) shows, $La$ can be interpreted as a characteristic value of 
the surface tension force times inertia 
divided by the square of the viscous force. 

A fourth dimensionless group that we shall have occasion to use is the 
Weber number 
\be
We = \frac{\rho a U^2}{\gamma}
\ee
which measures the ratio of inertia to surface tension. 
It is not independent of the three groups defined previously
because $We = Re^2/La$.

\section{Laboratory experiments}

\subsection{Experimental setups}

We used two experimental setups. The first (Setup 1) is the teaching
setup built by Plint \& Partners Ltd. (Deltalab) for the analysis of Poiseuille flow  
that we mentioned in the Introduction. The working fluid
is mineral oil (Total Azolla ZS22) with dynamic viscosity $\eta = 0.039$ Pa s,
density $\rho = 860$ kg m$^{-3}$, 
and surface tension coefficient
$\gamma = 0.037$ N m$^{-1}$. 

In order to be able to change the working fluid and pipe radius and
to make local measurements of the curtain thickness, we built a 
second setup (Setup 2). This involved a 1.2 m long pipe with
inner radius $a = 5.5$ or 8.5 mm, through which fluid is forced
by a centrifugal pump. The jet/curtain falls into a large tank whence
it is pumped back to the reservoir supplying the pipe. The flow rate
is controlled by adjusting the frequency of the pump. The working
fluids were water/glycerine mixtures, water/glucose syrup mixtures,
and silicone oil. At 25$^\circ$C the silicone oil
had $\rho = 960$ kg m$^{-3}$ (decreasing with temperature
at a rate of 1 kg m$^{-3}$ per K)
and $\gamma =  0.0207$ N m$^{-1}$.
Its kinematic viscosity was about $3\times 10^{-5}$ m$^2$ s$^{-1}$
and decreases with temperature by about 1\% per K.
Viscosities were measured with an Anton Paar MCR 501 rheometer
and surface tension with a Kruss DSA30 tensiometer.

Each of the two experimental setups has shortcomings that should be
kept in mind. The main disadvantage of Setup 1 is that it cannot
be disassembled, making it impossible to change the pipe 
or the working fluid. Moreover, the fluid is ejected into a closed chamber
that prevents access to the curtain. In Setup 2, there is a small
(amplitude 0.1 mm) oscillation of the pipe transmitted from the 
pump through the hoses. When the curtain is very thin, the oscillation
leads to the formation of bubbly surfaces alternately on the two sides.
A second disadvantage is that the temperature of the liquid increases
as passes repeatedly through the pump, changing its viscosity 
and modifying the flow rate. Finally, the flow rate is not perfectly 
constant on shorter time scales, and oscillates with a typical frequency $< 1$ Hz. 
Consequently, the curtain moves a little in its own plane, typically
by about 1 cm in $x$. 

\subsection{Observations}

\begin{figure}
\includegraphics[width=1\linewidth]{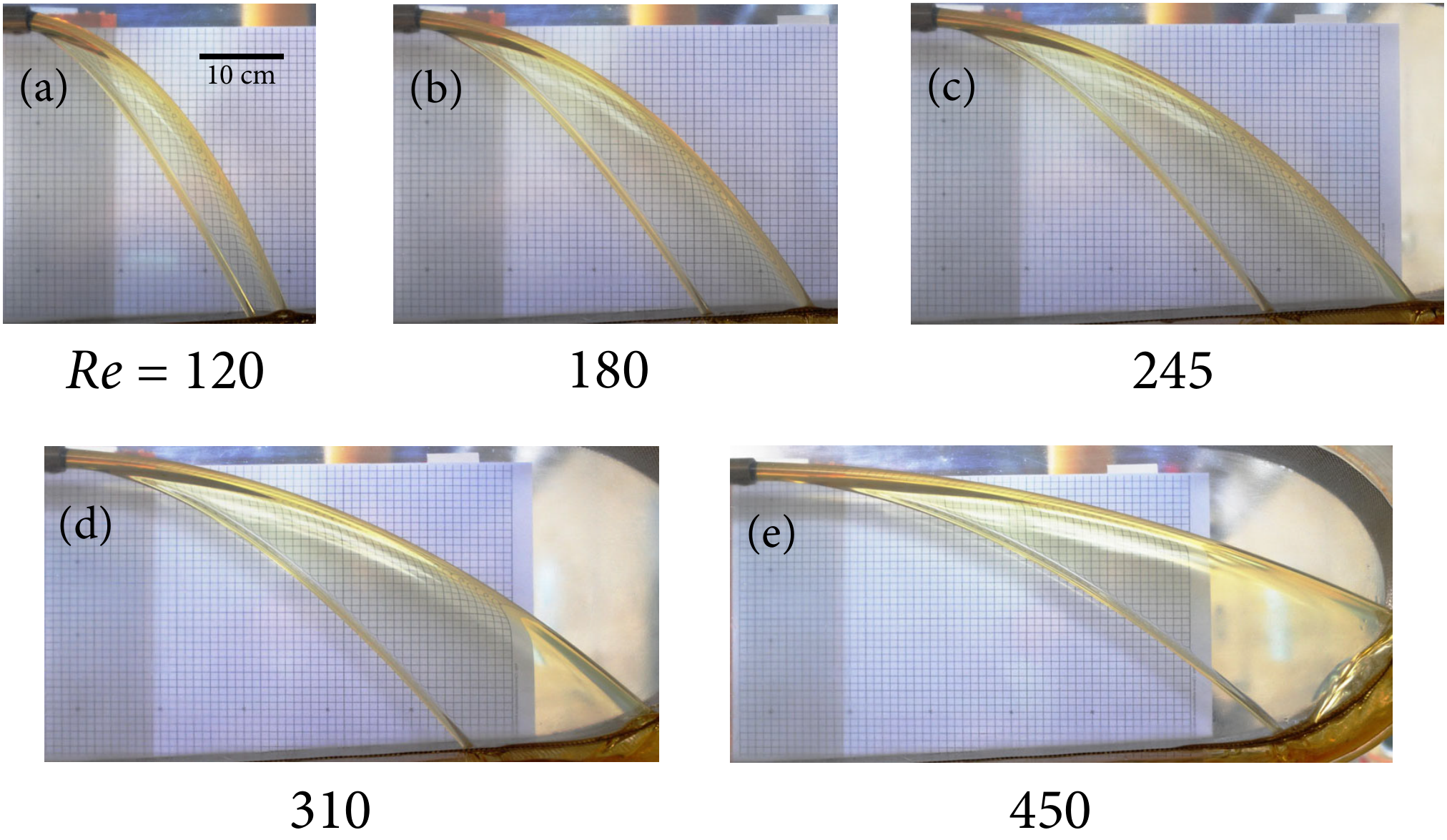}
\caption{\label{fig_shape_vs_re}
Shape of the curtain observed in Setup 1 as a function of the Reynolds number $Re=$ (a) 120,
(b) 180, (c) 245, (d) 310 and (e) 450.
The approximate values of the Dean and Laplace numbers are
$De = 58$ and $La = 160$.  The grid squares
in the background are 1 cm across. 
}
\end{figure}

We begin by using Setup 1 to examine how the morphology of the curtain
depends on the Reynolds number $Re$. Fig. \ref{fig_shape_vs_re} shows
the shape of the curtain for five values of $Re$. All five experiments
have $De = 58$ and $La = 160$. As $Re$ increases, both the primary and secondary jets
become more horizontal, as one would expect from purely ballistic
considerations. For $Re\geq 180$, the primary and
secondary jets move progressively further away from each other 
with increasing distance downstream. For $Re = 120$ (fig. \ref{fig_shape_vs_re}a), by contrast,
the two jets first diverge and then begin to approach each other. 

\begin{figure}
\includegraphics[width=0.8\linewidth]{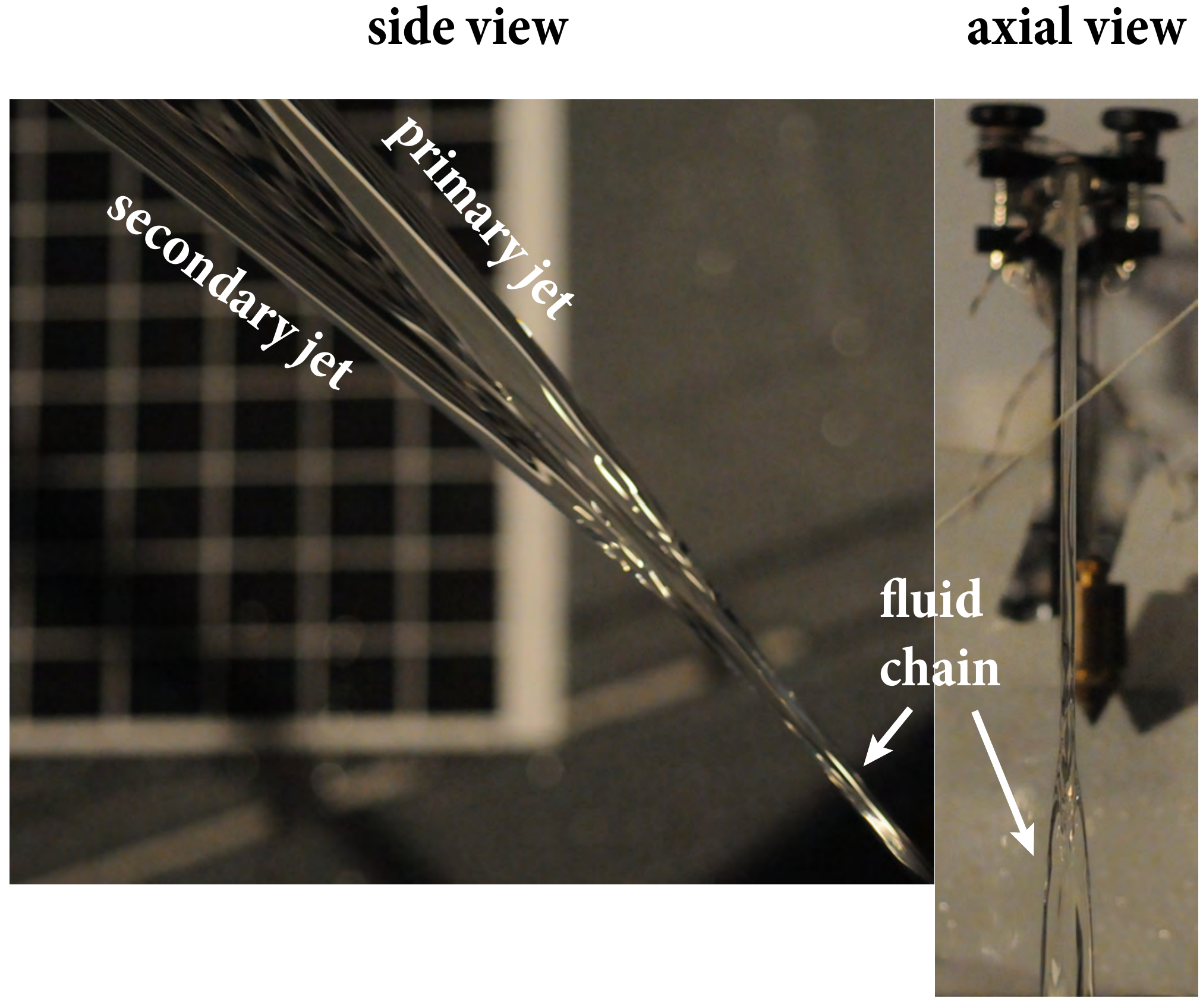}
\caption{\label{fig_collision}
Fluid chain resulting from the collision of the primary and secondary jets.
The experiment is performed using Setup 2 with $a= 5.5$ mm and diluted glucose syrup
as the working fluid.
The values of the dimensionless groups are 
$Re = 1360$, $De = 160$ and $La = 4300$. The grid squares
in the background are 2 cm across. 
}
\end{figure}

The approach of the primary and secondary jets in fig. \ref{fig_shape_vs_re}a
raises the question of what would happen further downstream if
the jets had not been blocked by the bottom of the transparent chamber. 
This can be explored using experiments with setup 2 and diluted glucose syrup,
for which the large surface tension effect pulls the secondary jet strongly
upwards towards the primary one. Fig. \ref{fig_collision} shows a 
close-up view of one such experiment with $Re = 1360$, $De = 160$ and $La = 4300$.
The approaching primary and secondary jets eventually collide to
form a `fluid chain' \citep{bush04jfm}. 

At large flow rate and for $a=8.5$ mm in Setup 2 the film grows quite thin far downstream from the pipe exit and 
becomes unstable. Sinuous waves of rapidly growing amplitude are observed, reminiscent of the flutter instability 
studied recently by Dighe and Gadgil\cite{dighe21jfm}. Further study of this instability is beyond the scope of this paper.

\subsubsection{Rescaling of the curtain shape}

\begin{figure}
\includegraphics[width=0.8\linewidth]{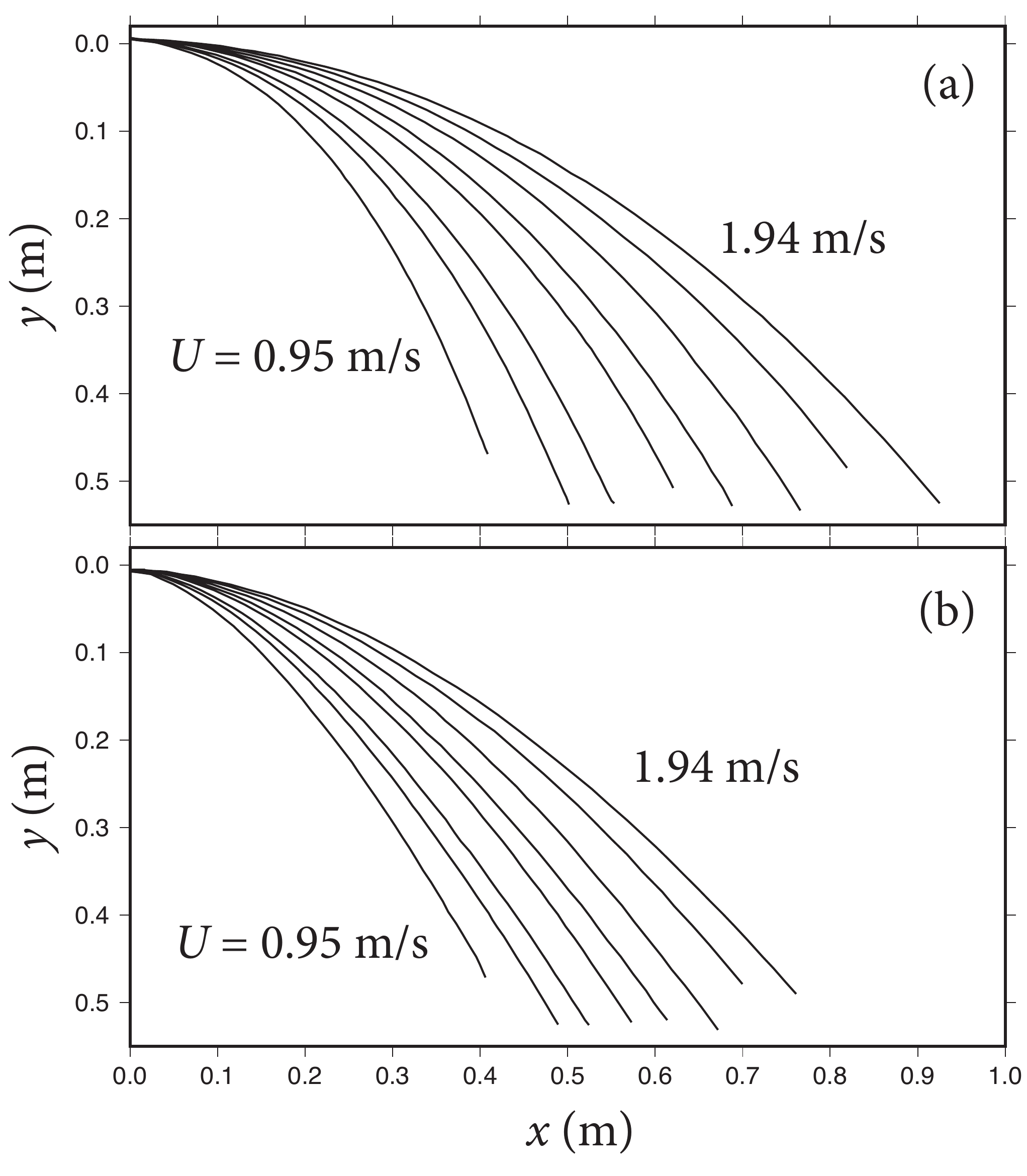}
\caption{\label{fig_5pt5_unscaled}
Profiles of (a) the upper boundary of the primary jet and (b) the lower boundary
of the secondary jet for different values
of the mean exit velocity $U$, obtained using Setup 2 with silicone
oil and a pipe having $a = 5.5$ mm. The corresponding dimensionless
parameters are $Re\in [160,330]$, $De = 40$ and $La = 120$.
}
\end{figure}

Next, we ask whether the trajectories of the primary and secondary jets
can be rescaled to yield universal curves. Fig. \ref{fig_5pt5_unscaled}a
shows 8 profiles of the upper boundary (top) and the lower boundary
(bottom) of the primary jet for different values
of the mean exit velocity $U$, obtained using Setup 2 with silicone
oil and a pipe having $a = 5.5$ mm. For these experiments $Re\in [160,330]$,
$De = 40$ and $La = 120$. 

\begin{figure}
\includegraphics[width=0.8\linewidth]{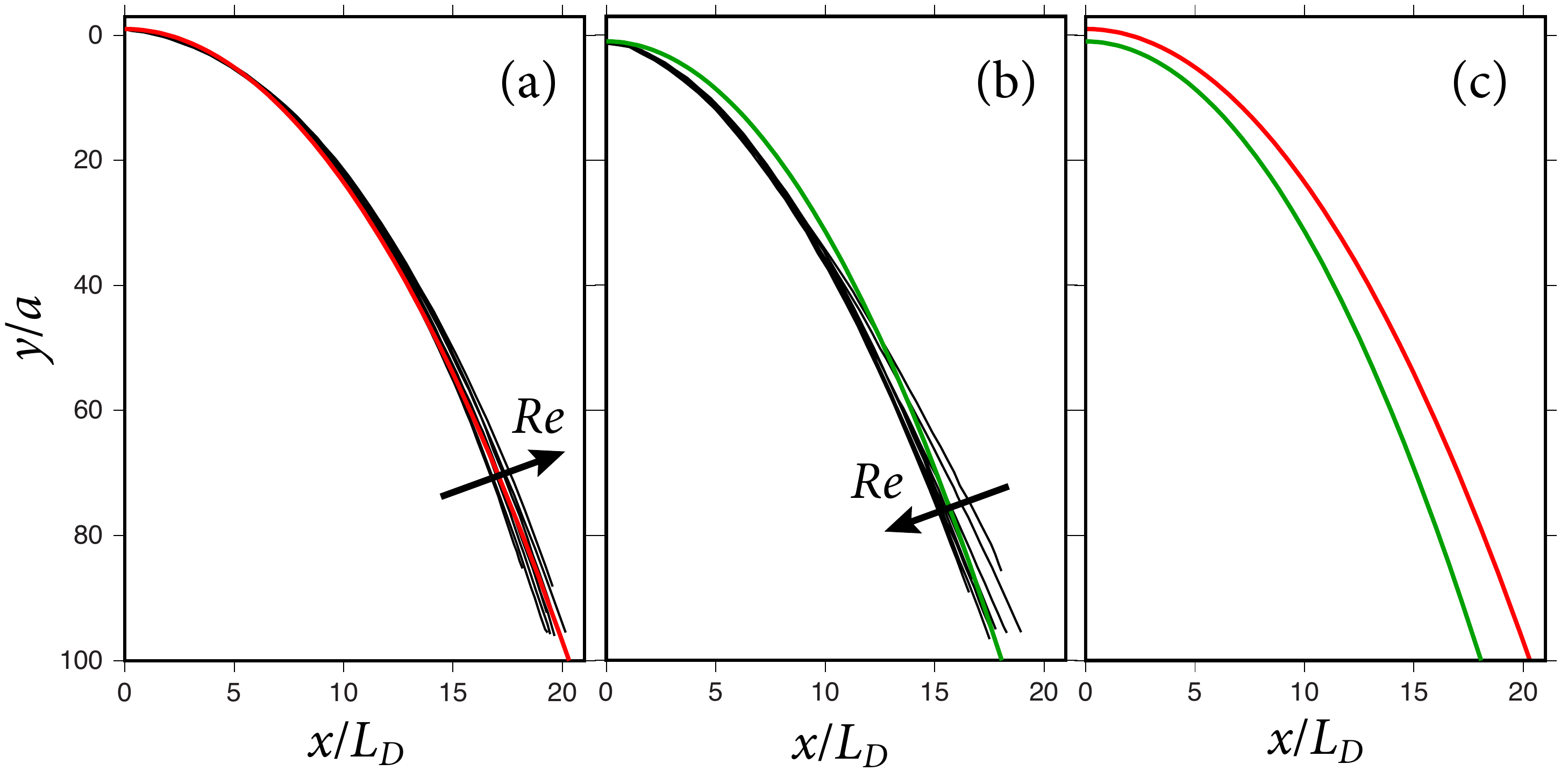}
\caption{\label{fig_5pt5_scaled}
Rescaled curves from fig. \ref{fig_5pt5_unscaled}. Parts a) and b) 
correspond to parts a) and b) of fig. \ref{fig_5pt5_unscaled}.
The red and green curves are parabolas that  best fit the experimental
data, and correspond to $U_{eff}/U$ = 1.43 and 1.28, respectively. The
arrows show the sense of increasing Reynolds number $Re$. Part (c) shows
the two parabolas together to facilitate comparison.
}
\end{figure}

To rescale the curves in fig. \ref{fig_5pt5_unscaled}, we choose new scaled
variables $Y = y/a$, $X = x/L_D$, where $L_D$ is the Dean length
defined by (\ref{deanlength}). Fig. \ref{fig_5pt5_scaled} shows the rescaled versions
of the curves of fig. \ref{fig_5pt5_unscaled}. All the data collapse 
onto a universal curve for $X < 10$.
For larger values of $X$ significant differences among the scaled
curves are evident, especially for the lower rim (fig. \ref{fig_5pt5_scaled}b). 

Proceeding in a Torricellian way, we now determine the
parabola that best fits the curves in 
fig. \ref{fig_5pt5_scaled}. 
The equation for the parabolic 
trajectory of a particle having an effective horizontal velocity $U_{eff}$ and
initial position $y=\mp a$ is
\be
y = \mp a + \frac{g x^2}{2 U^2_{eff}}\quad\to \quad Y  = \mp 1 + \frac{1}{2}\left(\frac{U}{U_{eff}}\right)^2 X^2,
\label{parab}
\ee
where 
the plus and minus signs correspond to the top and bottom of the jet at the pipe exit.
Using a simple least-squares fit to the data in fig. \ref{fig_5pt5_scaled} we obtain
the parabolas shown by the red and green lines in that figure.  The parabolic
fit to the data is good for the upper boundary (fig. \ref{fig_5pt5_scaled}a) but less
good for the lower boundary (fig. \ref{fig_5pt5_scaled}b).
The best-fitting parabolas for the upper and lower boundaries
have $U_{eff}/U = 1.43$ and  1.28, respectively. 
Note that $U_{eff}/U$ must be less than 2 because the fastest particles on the pipe axis have velocity 
$2U$.

\begin{figure}
\includegraphics[width=0.8\linewidth]{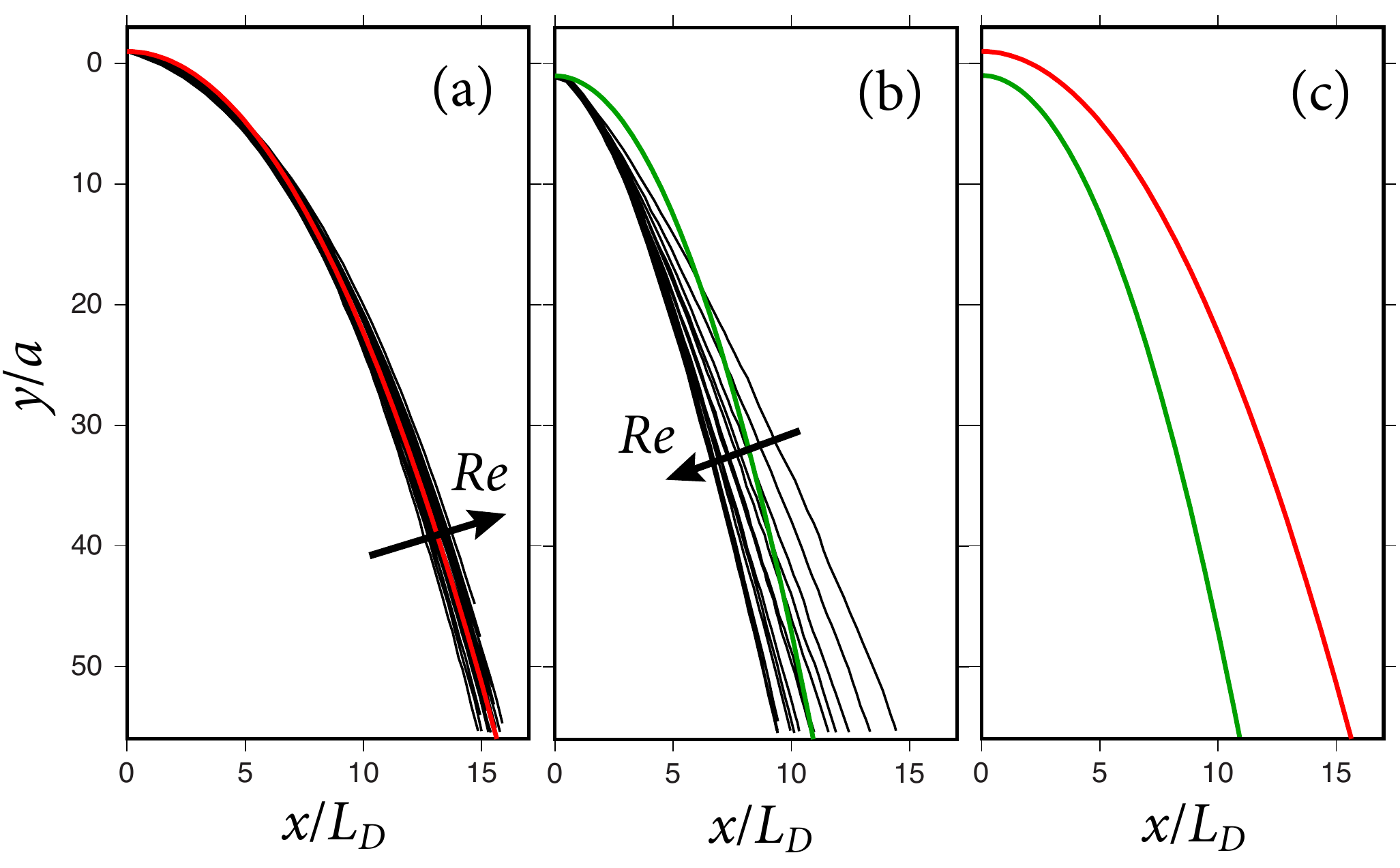}
\caption{\label{fig_8pt5_scaled}
Same as parts (a)-(c) of fig. \ref{fig_5pt5_scaled}, but for 17 experiments with
$a = 8.5$ mm. The corresponding dimensionless parameters are
$Re\in [150, 480]$, $De = 80$ and $La = 200$.  The best-fitting
parabolas (red and green lines) have $U_{eff}/U = 1.46$ and
1.04, respectively.
}
\end{figure}

Fig. \ref{fig_8pt5_scaled} is the same as fig. \ref{fig_5pt5_scaled}
but for a suite of 17 experiments with a larger pipe radius
$a = 8.5$ mm. The upper boundary of the primary jet (fig. \ref{fig_8pt5_scaled}a) is again well fit
by a parabola with $U_{eff}/U = 1.46$, nearly the same value as for the 
parabola in fig. \ref{fig_5pt5_scaled}a. By contrast, the data for the lower boundary
of the secondary jet (fig. \ref{fig_8pt5_scaled}b) clearly do not collapse onto a universal curve,
and are consequently poorly fit by a parabola (green line corresponding to 
$U_{eff}/U = 1.04$) that is shown only for
completeness. The secondary jet is clearly strongly affected by $Re$, being 
lower (in this dimensionless representation) when $Re$ is larger.

\subsubsection{Thickness measurements}

The thickness of the curtain in the $z$-direction (into the plane of the 
page) is measured in setup 2 using a local optical probe (Chromapoint from STIL, 
\url{http://point.stil-sensors.com/?lang=EN}). Incident white light traverses a chromatic lens and the reflected light 
is analysed. The color of the reflected light gives the distance to the interface. With a transparent liquid two resolved peaks can be observed in the spectrogram and, if the optical index of the liquid is known, the film thickness $h(x, y)$
can be measured with a 10 $\mu$m resolution and at a 1 kHz acquisition frequency.
The optical probe can be translated vertically or horizontally by an $x$-$y$ carriage. Unfortunately, thicknesses in the two jets cannot be measured with this method as the slopes of the interfaces are too large there. We note that the background grids visible in figures \ref{fig_structure},
\ref{fig_shape_vs_re}, \ref{fig_collision} and \ref{fig_trajectories}
are solely for calibration of distances in the $x$- and $y$-directions, and play
no role in the measurements of the curtain thickness.

\begin{figure}
\includegraphics[width=0.8\linewidth]{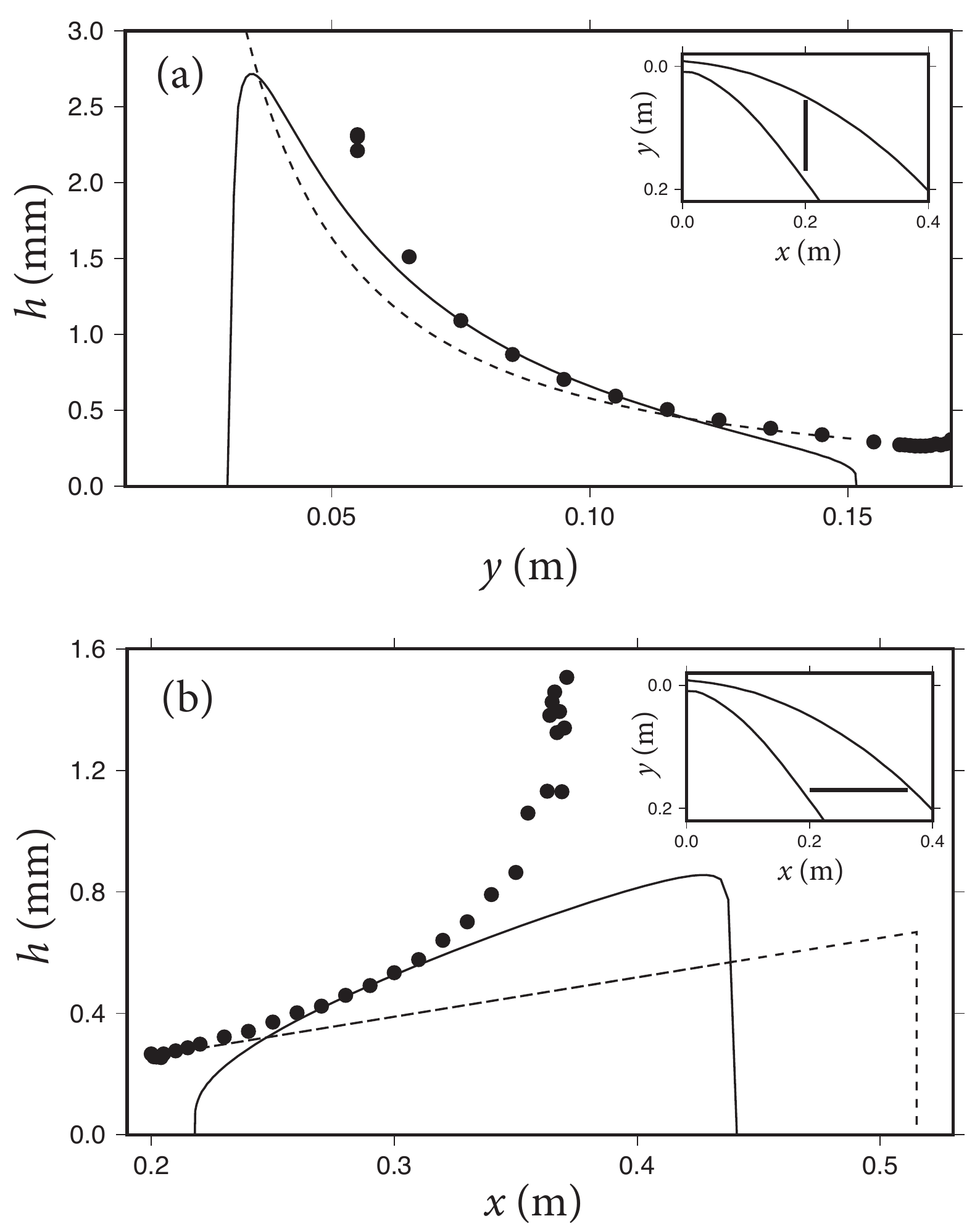}
\caption{\label{fig_hprofiles}
Black dots: vertical (a) and horizontal (b) profiles of the curtain thickness $h$ measured in a laboratory
experiment with $Re = 380$, $De=80$ and $La = 194$. The locations of the profiles
are shown in the insets. Dashed and solid lines are the predictions (\ref{hxyinviscid}) and (\ref{hinterm}) of the simple and
corrected ballistic models, respectively.
}
\end{figure}

The black dots in Figure~\ref{fig_hprofiles} show (a) vertical and (b) horizontal profiles of the thickness measured in a laboratory
experiment with $Re = 380$, $De=80$ and $La = 194$.  The thickness of the sheet 
in between the two jets increases in $x$ and decreases in $y$. 
The solid and dashed lines show theoretical predictions that
will be discussed later. 

\subsubsection{Dean recirculation}

Using setup 2, we were able to inject dye into the upper side of the jet close to its exit from the pipe.
The dye was observed by eye to follow spiral trajectories down the jet, corresponding to 
a secondary flow comprising two cells having longitudinal vorticity of opposite sign. 
Such behavior is reminiscent of Dean recirculation in curved pipes \citep{berger83annrev}.
We will have more to say about Dean recirculation below in \S~\ref{sec:dean2}.

\section{Theoretical analysis}

\subsection{Simple ballistic model}

\begin{figure}
\includegraphics[width=0.80\linewidth]{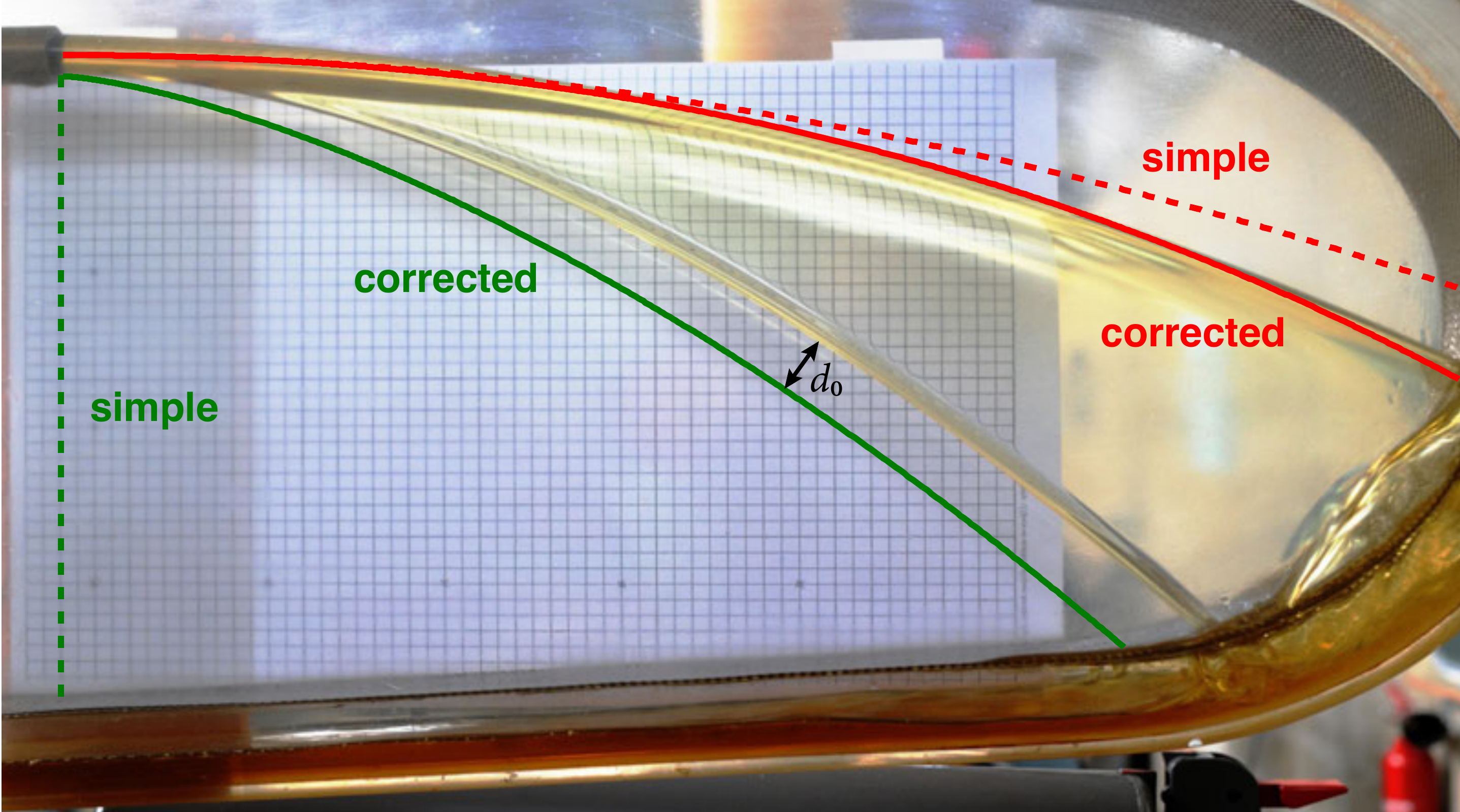}
\caption{\label{fig_trajectories}
Trajectories predicted by the simple (dashed lines) and corrected (solid lines) ballistic models,
compared with an experiment 
with $Re = 450$, $De = 64$ and $La = 200$. Red lines are trajectories of particles exiting the pipe
at the center $y=0$, and green lines are for particles exiting at the bottom $y = a$.
The black double-headed arrow indicates a typical difference $d_0$ between 
the prediction of the corrected ballistic model and the observed lower
extremity of the secondary jet. 
}
\end{figure}

To start simply, we present a zeroth-order model for the curtain by
supposing that fluid particles exiting the pipe do not interact with 
one another in any way. This amounts to ignoring viscosity and surface tension. 
Each particle therefore describes a parabolic
trajectory corresponding to its initial velocity of exit from the pipe.
The exit velocity $u_x(r)$ as a function of the radius $r$ across the pipe is that of a
developed Poiseuille flow, and is
\be
u_x(r) = U f_0(r),
\quad
f_0(r) = 2\left( 1 - \frac{r^2}{a^2}\right).
\label{poisic}
\ee
The equations governing the trajectories are the ballistic equations
$\dot x = u_x(r)$ and $\ddot y = g$, where dots denote derivatives 
with respect to the time $t$. Solving these and eliminating the 
parameter $t$, we obtain
\be
y = r + \frac{ g x^2}{2 u_x(r)^2}.
\label{trajinv}
\ee
The red and green dashed lines in Figure \ref{fig_trajectories} show the trajectories (\ref{trajinv}) 
with $r = 0$ (red) and $r=a$ (green), for an experiment using Setup 1.
The red dashed line is a parabola corresponding to the maximum
exit velocity $u_x(0) = 2 U$. It follows reasonably closely the trajectory of the 
primary jet, but is systematically too high. The vertical green dashed line is the (degenerate)
parabolic trajectory corresponding to the minimum exit velocity $u_x(a) = 0$.
It bears no relation to the trajectory of the secondary jet, indicating that a 
simple ballistic model is not valid for particles exiting the pipe near its wall. 

The expression (\ref{trajinv}) for the trajectories allow us to derive an expression
for the thickness $h(x,y)$ of the curtain. We sketch the derivation here to illustrate
the basic idea, which will be applied to a more realistic model in the next section. 
For simplicity, we neglect the finite radius of the pipe, which is equivalent to 
making a `far-field' assumption $(x,y)\gg a$. The equation (\ref{trajinv}) for the trajectories
now becomes
\be
y(x,r) = \frac{ g x^2}{2 u_x(r)^2}.
\label{trajinvff}
\ee
Denoting by $h_1$ the thickness predicted by the simple ballistic model, we equate the flux of fluid exiting the pipe through
an annulus of area $2\pi r \mathrm{d} r$ to the flux crossing a vertical plane
at some fixed downstream position $x$. This requires 
\be
2\pi r u_x(r) \mathrm d r = h_1 u_x(r)\mathrm d y,
\label{fluxbalance1}
\ee
where $\mathrm d y$ is the infinitesimal height of the portion of the curtain through
which fluid from the annulus of width $\mathrm d r$ passes. Rearranging
(\ref{fluxbalance1}), we obtain
\be
h_1 = 2\pi r\left( \frac{\partial y}{\partial r}\right)^{-1}.
\label{hinv1}
\ee
Differentiating (\ref{trajinvff}) with respect to $r$ and 
substituting the result into  (\ref{hinv1}),
we obtain
\be
h_1(x,y) = \frac{\pi a^2}{4 U}\left(\frac{g}{2 y^3}\right)^{1/2} x.
\label{hxyinviscid}
\ee
Introducing $H_1=h_1/a$ the foregoing equation can be written in dimensionless form as
\be
H_1(X,Y) = \frac{\pi}{4 \sqrt{2}} \, Y^{-3/2}X.
\label{hxyinviscid_dimensionless}
\ee
Eqn. (\ref{hxyinviscid_dimensionless}) predicts that for constant $Y$ the curtain
has a thin `wedge' shape with $H_1 \propto X$. We shall compare the prediction
(\ref{hxyinviscid}) with observations after introducing a corrected ballistic model
that accounts for the effects of viscosity. 

\subsection{Corrected ballistic model}

In this section we propose a corrected ballistic model for the morphology of the jet.
The idea is to model the flow as the composition of two components: the axisymmetric
flow within a steady horizontal jet in the zero-gravity limit, and a downward vertical velocity
due to free fall under gravity. The model differs from the simple ballistic model
of the previous section by accounting for the effects of viscosity 
on the axisymmetric horizontal jet. 

\begin{figure}
\includegraphics[width=0.70\linewidth]{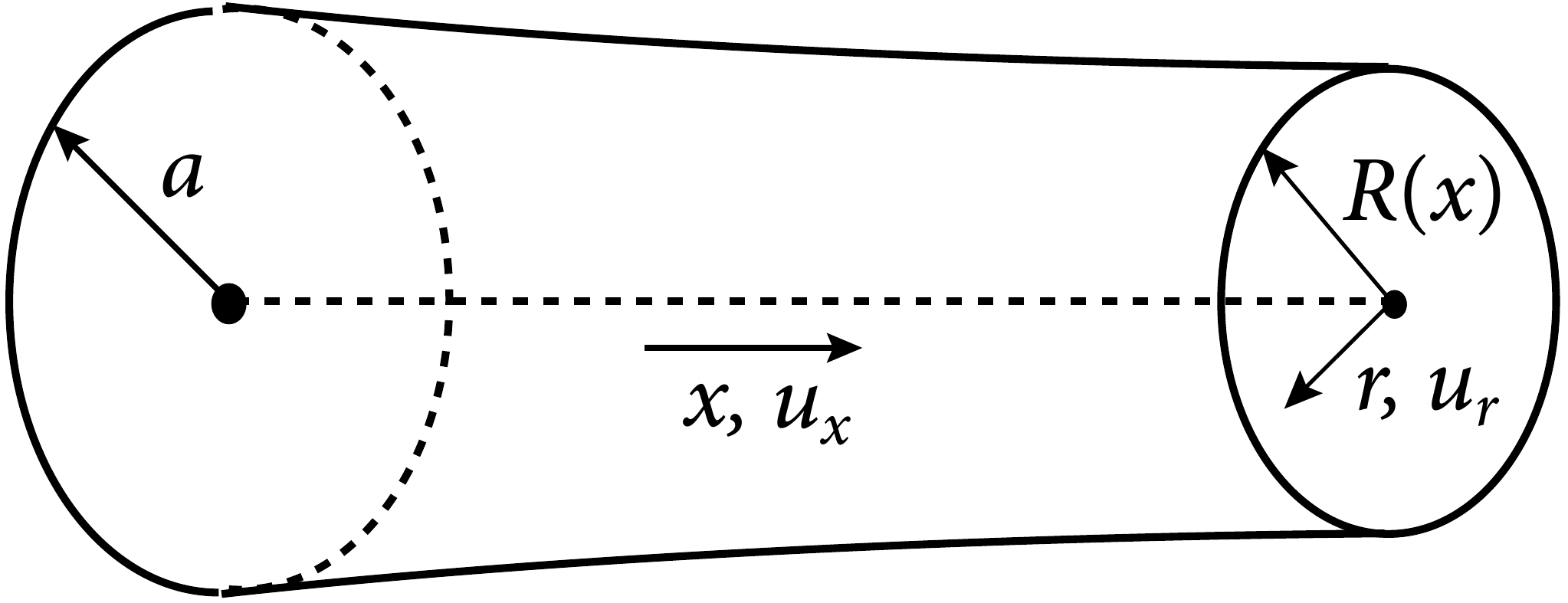}
\caption{\label{fig_defsketch}
Definition sketch of a steady axisymmetric horizontal jet in the absence of gravity.
The axial and radial components of the velocity are $u_x$ and $u_r$, respectively.
The radius of the nozzle is $a$, and the radius of the jet is $R(x)$, where
$x$ is the distance from the nozzle. 
}
\end{figure}

The first task is to determine the steady axisymmetric flow in a horizontal jet in the absence of gravity. 
Figure \ref{fig_defsketch} shows a definition sketch of the jet.
The velocity within the jet is $\mathbf u = u_r\mathbf e_r + u_x \mathbf e_x$, where
$\mathbf e_r$ and $\mathbf e_x$ are unit vectors in the directions indicated by
subscripts. The radius of the jet is $R(x)$, and the radius of the pipe is $a$. 

Following Sevilla\cite{sevilla11},
the governing equations in the boundary-layer approximation are
\be
r \frac{\partial u_x}{\partial x} + \frac{\partial}{\partial r}(r u_r) = 0,
\label{continuity}
\ee
\be
u_x \frac{\partial u_x}{\partial x} + u_r \frac{\partial u_x}{\partial r} = \frac{\gamma R'}{\rho R^2} + \frac{\nu}{r}\frac{\partial}{\partial r}\left( r\frac{\partial u_x}{\partial r}\right)
\label{axialmom}
\ee
where $\gamma$ is the coefficient of surface tension and $R' = \mathrm d R/\mathrm d x$. 
Eqn. (\ref{continuity}) is the 
continuity equation, and (\ref{axialmom}) is the 
axial momentum equation. The radial momentum equation in the boundary-layer approximation
simply states that the pressure does not vary across the jet. This allows
one to set the pressure everywhere equal to the capillary pressure $p_c$.
Because $R(x)$ is assumed to vary slowly over a length scale $a Re$, 
the capillary pressure gradient $-\mathrm d p_c/\mathrm d x \equiv \gamma R'/\rho R^2$
is due only to the azimuthal curvature of (circular) jet cross-sections, and does not take into
account the second (and much smaller) principal curvature in the axial direction.

Equations (\ref{continuity}) and (\ref{axialmom}) must be solved subject to the conditions
\be
u_x(0,r) = U f_0(r),
\quad    f_0 =  2 (1 - r^2/a^2),
\label{poiseuille}
\ee
\be
\frac{\partial u_x}{\partial r}(x, a) = 0,
\label{freesurface}
\ee
\be
u_r (x,0) = \frac{\partial u_x}{\partial r}(x,0) = 0,
\label{axialbc}
\ee
\be
u_r(x,a) = R' u_x(x,a).
\label{kincond}
\ee
Condition (\ref{poiseuille}) states that the flow exiting the nozzle $x=0$ has 
a parabolic (Poiseuille) profile with a mean exit velocity $U$.
Condition (\ref{freesurface}) states that the outer surface of the jet is
free of shear traction. Conditions (\ref{axialbc}) state that the radial
velocity and the radial derivative of the axial velocity vanish on the 
axis $r=0$ of the jet. 
Finally, condition (\ref{kincond}) is the kinematic condition on the 
jet's outer surface. 

\begin{figure}
\includegraphics[width=0.80\linewidth]{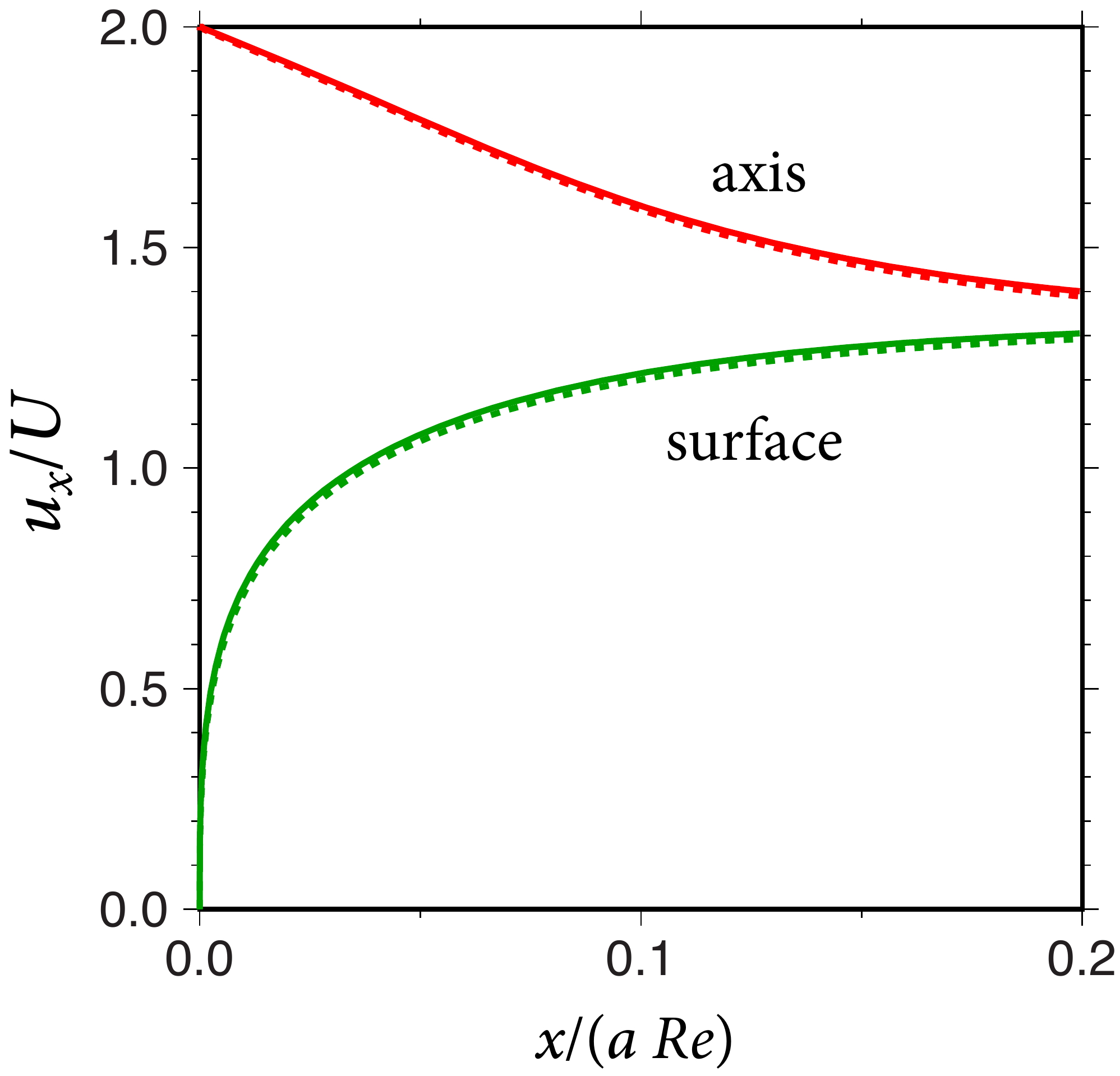}
\caption{\label{fig_horizjet_uvsx}
Velocities $u_x(x)$ on the axis (red lines) and the outer surface (green lines) of 
an axisymmetric horizontal jet in the absence of gravity, for
$We = 10$ (dashed lines) and 1000 (solid lines).
}
\end{figure}

We now nondimensionalize the foregoing boundary-value problem
by introducing new dimensionless variables $r/a$, $x/(a Re)$, $u_x/U$, and $u_r Re/U$. 
The only two dimensionless groups that appear are then $Re$ and $We$.
We solved the  dimensionless problem
numerically using the method of lines outlined by 
Gordillo \textit{et al.}\cite{gordillo01}. 
Of primary interest for our corrected ballistic model are the axial velocities
$u_x(x,0)$ and $u_x(x,a)$ along streamlines corresponding 
respectively to the central axis $r=0$ and to the outer surface $r=a$ of the jet. These are shown
in Figure \ref{fig_horizjet_uvsx} for $We = 10$ (dashed lines) and 1000 (solid lines). Fluid on the outer surface accelerates
rapidly, and fluid on the axis decelerates more slowly, both tending towards the same
final plug flow velocity $(4/3) U$. This behavior is due to radial viscous diffusion of 
vorticity consequent upon the sudden change in the outer boundary condition
from rigid to free when the jet exits the pipe. The dashed and solid curves in 
Figure \ref{fig_horizjet_uvsx} are almost identical, indicating that surface
tension plays only a minor role in the dynamics of horizontal jets when $We \gg 1$. 

\subsubsection{Particle trajectories}

The horizontal velocities shown in Figure \ref{fig_horizjet_uvsx} can now be transformed 
into nominal particle trajectories by adding a vertical component of motion
corresponding to free fall. Let $y$ be the downward vertical coordinate with
origin at the center of the pipe. The horizontal velocity of the axisymmetric jet is 
\be
u_x = U f\left(\zeta, r\right),
\quad
\zeta = \frac{x}{a Re}  
\label{horizcomp}
\ee
where $r\in [0, a]$ is the radial coordinate across the pipe. The function 
$f(\zeta, 0)$ is the upper curve in Figure \ref{fig_horizjet_uvsx},
and $f(\zeta, a)$ is the lower curve in that figure. 
We now combine (\ref{horizcomp}) with the free fall equation $\ddot y = g$,
obtaining
\be
y(x,r) = r + \frac{g}{2} 
\left[ \frac{a^2}{\nu} I(x,r) \right]^2,
\quad
I(x,r) = 
\int_0^{x/(a Re)}\frac{\mathrm d\zeta}{f(\zeta,r)}.
\label{trajectoriesvisc}
\ee
The dimensionless integral $I$ is just the time required for a particle
starting at a radius $r$ to travel
a distance $x$, measured in units of the characteristic viscous diffusion time $a^2/\nu$.

The red and green solid lines in Figure \ref{fig_trajectories} show
the trajectories predicted by the corrected ballistic model for the reference
experiment. The solid red line now follows closely the position of the 
primary jet, indicating that the corrected ballistic model
is accurate for that jet. The solid green line staring at $y=a$ is also much closer to the 
position of the secondary jet than was the prediction of the simple ballistic model (dashed green line). 
Nevertheless, the corrected ballistic model still predicts
that the lower extremity of the secondary jet lies substantially lower than its true
position. There are at least two possible reasons for this discrepancy.
One is that the model envisions purely radial viscous diffusion in an
axisymmetric jet, which is no longer a valid assumption when the 
lower part of the jet has become strongly non-axisymmetric. A second reason is the neglect
in the model of the surface tension-driven retraction
of the secondary jet and the attached thin fluid curtain, which
counteracts their tendency to move downwards under gravity.   
This effect will be estimated shortly in \S~\ref{sec_st}.

\subsubsection{Curtain thickness}

The corrected ballistic model allows us to predict the curtain's
thickness $h_2(x,y)$ as a function of position. Balancing fluxes as we
did in the previous section, we obtain
\be
2\pi r U f_0(r) \mathrm d r = u_x(x,r) h_2 \mathrm d y\quad \to \quad h_2(x,r) = 2\pi r \frac{f_0(r)}{f(\zeta, r)}
\left( \frac{\partial y}{\partial r}\right)^{-1}
\label{hinterm}
\ee 
where $y(x,r)$ is given by (\ref{trajectoriesvisc}). 
Equations (\ref{trajectoriesvisc}) and (\ref{hinterm}) are two parametric equations
for $y(x,r)$ and $h_2(x,r)$ with $r$ as a parameter, and 
must be solved numerically. It is easy to verify that (\ref{hinterm}) reduces 
to the corresponding inviscid result (\ref{hxyinviscid}) by setting 
$f(\zeta, r) = f_0(r)$ and making the far-field assumption by neglecting
the first term ($=r$) on the right-hand side of (\ref{trajectoriesvisc}). 

We are now in a position to compare the predictions $h_1(x,y)$ and $h_2(x,y)$
of our two ballistic models with the experimental measurements
shown by the black circles in fig. \ref{fig_hprofiles}. 
That figure also shows (a) vertical
and (b) horizontal profiles of $h_1$ (dashed lines) and $h_2$ (solid lines)
for the parameters of the experiment.
Both ballistic models do a reasonably good job of predicting the vertical
profile (fig. \ref{fig_hprofiles}b), including the rapid thickening for small values $y\approx 0.05$-0.06 m
as the primary jet is being approached. The situation for the horizontal
profile of fig. \ref{fig_hprofiles}b is more complicated. Each ballistic
model predicts well a portion of the observations at small values of 
$x < 0.32$. However, neither model predicts the rapid increase in 
thickness for larger values of $x$.

\subsubsection{Effect of surface tension}
\label{sec_st}

We now investigate whether the discrepancy between the observed
and calculated positions of the secondary jet (fig. \ref{fig_trajectories}) 
is due to the 
(hitherto unmodelled) effect of surface tension. 
To begin, we recall the expression derived by Taylor\cite{taylor59procrsoc}
and Culick\cite{culick60jap} for the velocity $U_{tc}$ of surface tension-driven
retraction of the free edge of a two-dimensional viscous sheet with
constant thickness $h$:
\be
U_{tc} = \left(\frac{2\gamma}{\rho h}\right)^{1/2}.
\label{vtc}
\ee
The formula (\ref{vtc}) translates the balance between
surface tension and inertia at the free edge. 
Refer now to fig. \ref{fig_trajectories}, in which the black double-headed arrow
indicates the typical distance $d_0$ between the 
numerically predicted and true lowermost streamlines of the secondary jet.  
The corrected ballistic model also predicts the travel time of a fluid particle from the 
pipe exit to any point on the green streamline; let $\tau_0$ be the
travel time to the point at the base of the black double-headed
arrow in fig. \ref{fig_trajectories}. We now ask: can the separation $d_0$ be 
accounted for by Taylor-Culick (T-C) retraction of the curtain's
free lower edge during the time $\tau_0$? Because we do not know the
appropriate value of $h$ to use, we solve the equation $d_0 = U_{tc} \tau_0$
for $h = h_{eff}$. We obtain:
\be
h_{eff} = \frac{2\gamma\tau_0^2}{\rho d_0^2}.
\label{heff}
\ee
The quantity $h_{eff}$ is the effective thickness of the curtain for which the
T-C formula (\ref{vtc})  predicts retraction by an amount $d_0$
in a time $\tau_0$. 
For the experiment of fig. \ref{fig_trajectories}, 
$d_0 = 0.032$ m and 
$\tau_0 = 0.20$ s, whence (\ref{heff}) predicts
$h_{eff} = 3.5$ mm.  This value of $h_{eff}$ is significantly larger than
the typical curtain thicknesses $h\approx 0.3$-1 mm shown in fig. \ref{fig_hprofiles}.
For such thicknesses, T-C retraction is too fast
(by a factor of $\approx 3$-4) to explain the observed separation $d_0$
in Figure \ref{fig_trajectories}. However, the foregoing calculation is
subject to significant uncertainty given that the T-C formula applies to an 
idealized inviscid two-dimensional sheet of constant thickness. Moreover, we note
that in our experiments with sugar syrup, the secondary jet is strongly
``pulled up" towards the primary jet, an effect that is certainly due to
the large surface tension of the syrup. This observation leads us to 
conclude that surface tension plays an important role in determining
the trajectory of the secondary jet, both for sugar syrup and (to a lesser
extent) for the mineral oil used in the experiment of Figure \ref{fig_trajectories}.

\section{Direct numerical simulation}
\label{dns}

To gain further insight into the 
morphology and dynamics of Torricelli's curtain, we performed a direct numerical simulation
using the volume-of-fluid code Gerris Flow Solver (GFS)
\citep{popinet03}. This code solves for the flow in a system
comprising two fluid phases, 
oil and the surrounding air in our case. Our flow domain is a rectangular
box, $0\leq x\leq 24 a$,
$-2a \geq y\geq 14 a$, and $0\leq z\leq 4a$, where the origin
is the center of the nozzle and $y$ increases downward. 
Because the jet has mirror symmetry across the vertical
$x$-$y$ plane, we solve the problem only for the half-domain
$z > 0$. On the left boundary $x=0$, we impose the
boundary condition (\ref{poisic}) corresponding to developed Poiseuille flow.
The two fluid phases are distinguished
by values of a phase variable $T = 0$ (air) and $T=1$ (oil), separated
by a thin diffuse interface across which $T$ varies rapidly from
0 to 1. 

Torricelli's curtain is quite challenging to simulate numerically due to the 
three-dimensionality of the problem and the
wide separation of lengthscales between the pipe diameter and the thinnest part
of the curtain.  It is therefore necessary
to exploit the octree adaptive grid refinement capacity of GFS, 
which divides cubes at a given refinement level into eight smaller cubes. 
The numerical flow domain initially consisted of 24 cubes with
sides equal to $4a$.  
We used a maximum grid refinement of a factor of $2^8 = 256$,
so that the smallest cube of the refined grid has a side
equal to $a/64$. At the end of our simulations, the grid comprised a maximum of
$\approx 1.6\times 10^6$ cubes. 

For our first simulation, we chose parameter values 
corresponding to one of our laboratory experiments with Setup 2 using silicone oil,
for which the values of the dimensionless groups are 
$Re = 480$, $De = 80$ and $La = 200$. 
We ran the simulation for a dimensional time $t = 0.63$ s, which corresponds
to 5.3 transit times across the length ($=24 a$) of the numerical box at the mean velocity $U$. 
We verified that this run time was
sufficient to achieve a steady state by comparing three solutions obtained using
a grid refinement of $2^7$ for times 0.33 s, 0.46 s, and 0.58 s. The shapes of
the jet at 0.33 s and 0.46 s were noticeably different, but those for
0.46 s and 0.58 s were indistinguishable. 

\subsection{Shape and thickness of the curtain}

\begin{figure}
\includegraphics[width=0.80\linewidth]{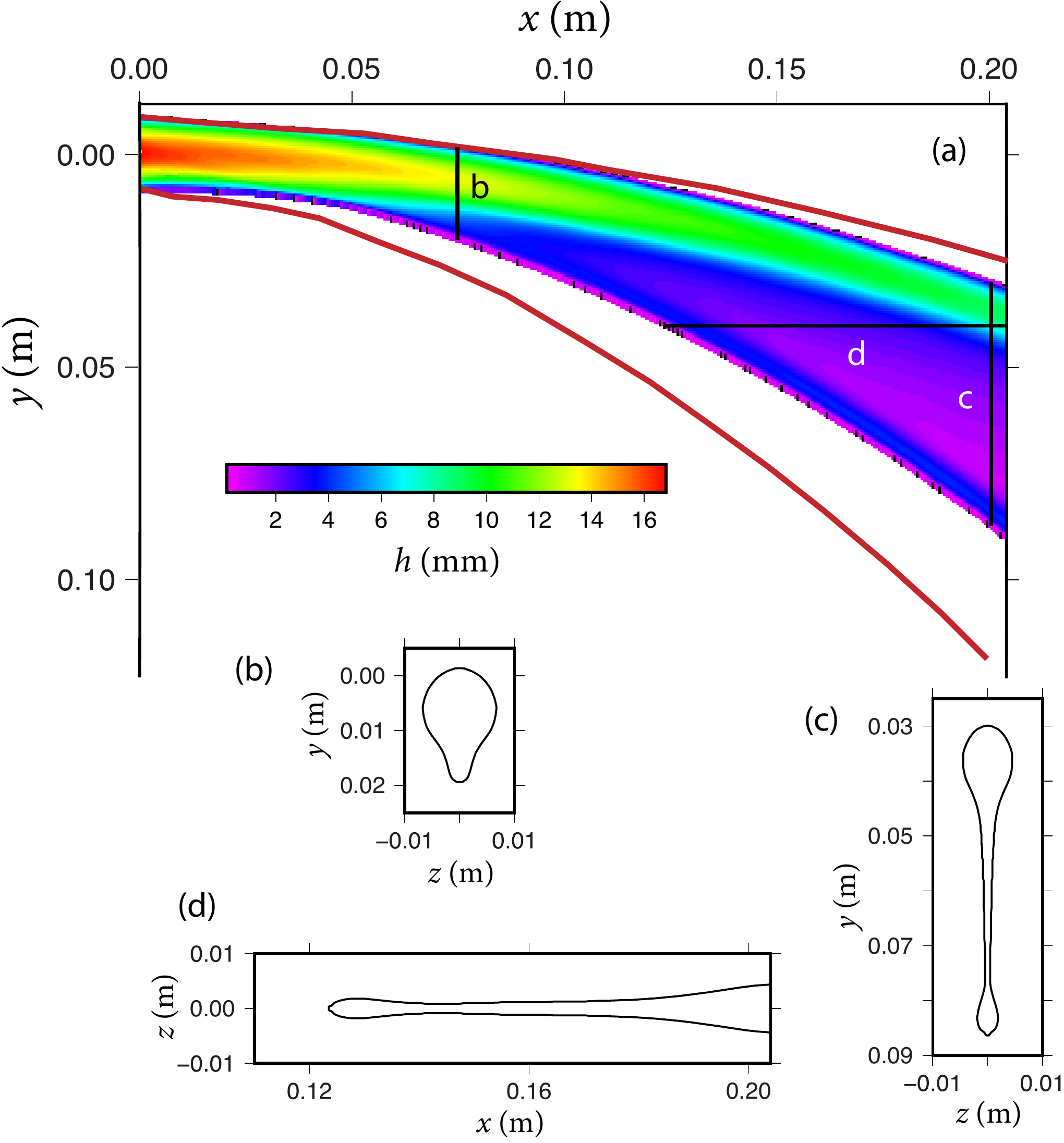}
\caption{\label{fig_silicone37_gerris_hxy}
(a) Thickness $h(x,y)$ of a curtain simulated using Gerris Flow Solver
\citep{popinet03}
with $Re = 480$, $De = 80$, and $La = 200$. 
Brick red lines are the experimentally observed extremities of the curtain for the same parameters. 
(b)-(d)  Vertical and horizontal cross-sections of the curtain along the black lines labelled b, c
and d in part (a).
}
\end{figure}

Figure \ref{fig_silicone37_gerris_hxy}a shows the thickness $h(x,y)$ of
the simulated curtain.
For comparison, the brick red lines show
the upper and lower extremities of the primary and secondary jets measured from a photograph of
the laboratory experiment. The agreement between the experiment and
the numerics is acceptable for the primary jet, but poor for
the secondary jet. Results
were similar for other simulations for parameters corresponding to other laboratory experiments.
In the Discussion below we speculate on the likely causes of the
discrepancy between the numerics and the experiments; briefly, we think
it is due to a difference in the effective upstream velocity boundary condition. 
However, we pursue the simulations here in order to obtain valuable information that is
not accessible in the laboratory experiments. 

We begin by examining some vertical and horizontal cross-sections of the curtain
shown in fig. \ref{fig_silicone37_gerris_hxy}a. The locations of the cross sections
are indicated by the vertical and horizontal black lines labelled b, c and d. 
The corresponding thickness profiles are shown in parts (b), (c) and (d) of the figure. 
After remaining nearly circular out to distances of a few pipe diameters, the 
cross-section of the curtain develops a pronounced up-down asymmetry, with
a nascent curtain beneath the primary jet (profile b). By the time the fluid has
reached a distance
$x = 0.2$ m, the initial asymmetry has developed 
into the now-familiar structure comprising primary and 
secondary jets connected by a thin curtain (profile c). The horizontal
profile in part (d) shows that the thickness of the curtain between the two jets 
increases approximately linearly with $x$, in qualitative agreement with the 
predictions of the two ballistic models.

\subsection{Particle trajectories and Dean recirculation}
\label{sec:dean2}

\begin{figure}
\includegraphics[width=0.80\linewidth]{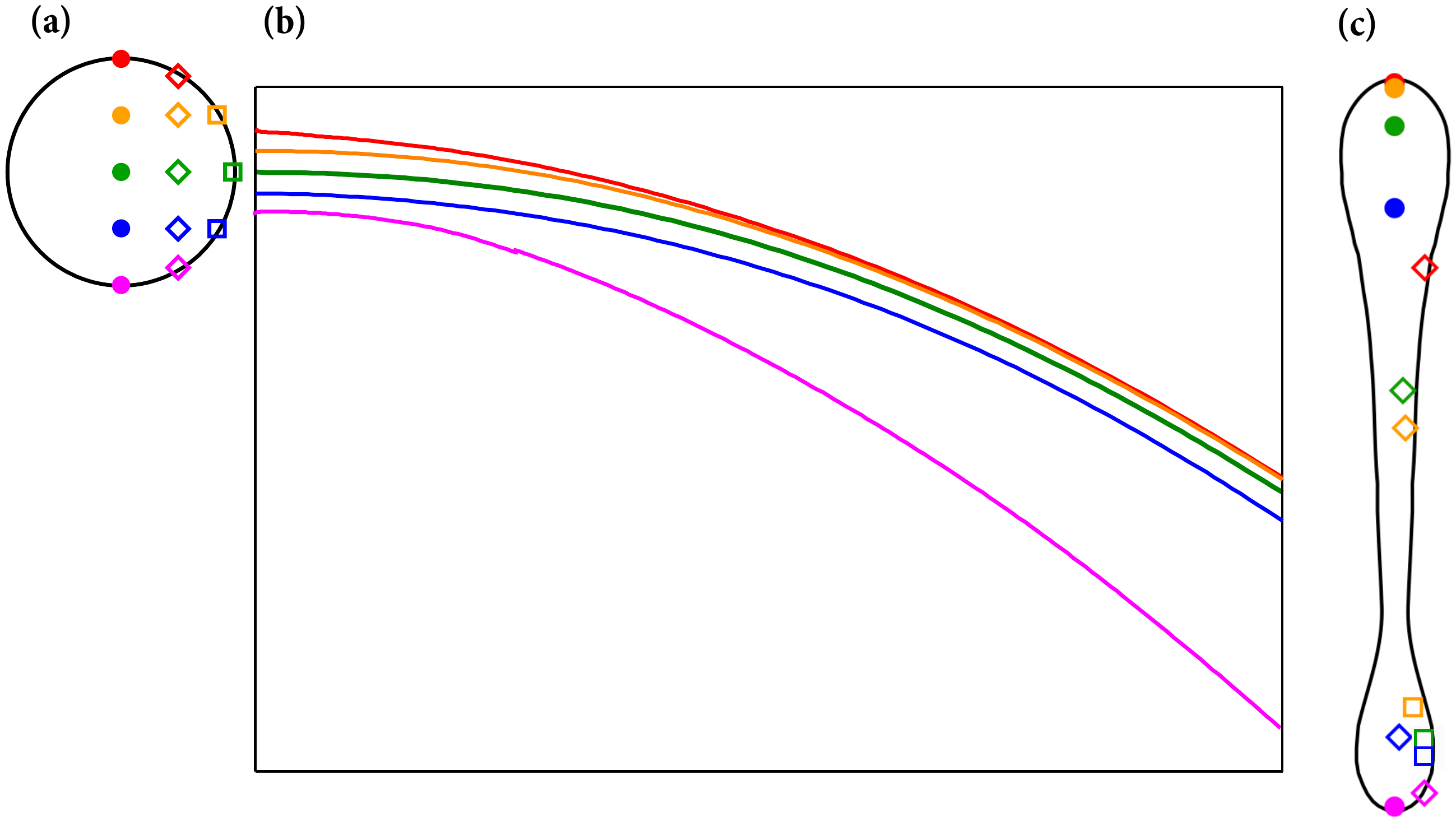}
\caption{\label{fig_streamlines}
Trajectories of selected material particles predicted by direct numerical simulation of 
a curtain with $Re = 245$, $De = 58$ and $La = 161$.  (a) Distribution of 13
material particles over the cross-section of the jet exiting the pipe. (b) Selected
trajectories in the symmetry plane $z=0$. (c) Positions of the 13 material particles
in the curtain at the distance $x=24a$ corresponding to the 
right boundary of part (b).
}
\end{figure}

The simulations also allow us to calculate three-dimensional particle trajectories within the curtain. Because these
are difficult to visualize, we instead examine the motion
of a handful of material particles. Fig. \ref{fig_streamlines}a illustrates the motion
of 13 material particles in a GFS numerical simulation with $Re = 245$, $De = 58$ and $La = 161$.
The particles are initially distributed uniformly over the circular cross-section of the 
jet as it exits the pipe. Fig. \ref{fig_streamlines}b shows five representative trajectories in the symmetry 
plane $z=0$ of the curtain; the progressive vertical widening of the curtain with distance
downstream is evident. Finally, Fig. \ref{fig_streamlines}c shows the positions of the 
13 material particles at the distance $x = 24 a$ corresponding to the 
right boundary of Fig. \ref{fig_streamlines}b. Note that the fluid in the secondary (lower)
jet originates mainly from the outermost layer of the jet exiting the pipe, which
flows downward around the circumference of the primary jet. 

\begin{figure}
\includegraphics[width=0.60\linewidth]{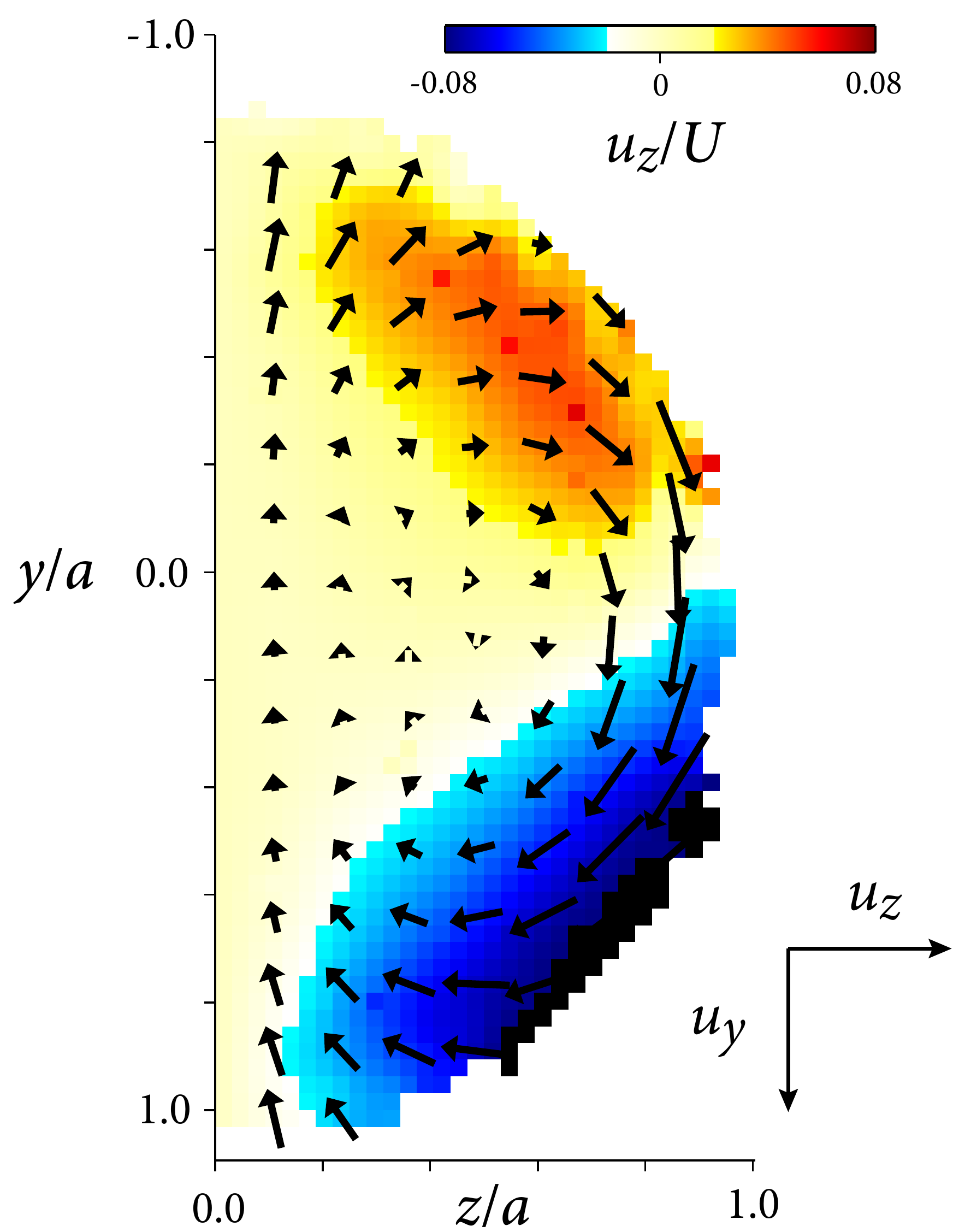}
\caption{\label{fig_recirculation}
Velocity field (projected onto the section) in a vertical plane $x = 2a$,
for the numerical simulation shown in fig. \ref{fig_streamlines}. The velocity field shown does 
not include the mean downward velocity of the section, which has been removed.
}
\end{figure}

Finally, we use our simulations to investigate the velocity field within
a vertical cross-section of the jet close to the pipe, where the cross-section
is still nearly circular. To make clearer the structure of the velocity field, we first
subtract from it the mean vertical (downward) velocity of the section. The 
result is shown in fig. \ref{fig_recirculation} for a section at $x = 2a$ in the
same numerical simulation shown in Fig. \ref{fig_streamlines}. The flow
consists of a recirculating vortex with downward flow on the outside of 
the jet, together with a mirror-image vortex on the other side $z<0$ of the jet. 
Even though we are at a relatively short distance ($=2a$) from the pipe exit, the amplitude of the 
horizontal velocity $u_z$ is already about $\pm 8\%$ of the mean exit
velocity $U$.

Recirculating vortices are prominent features of flow in curved pipes,
in which context they are known as Dean vortices \citep{berger83annrev}.
Such vortices are required by the conservation of angular momentum,
and can be understood in terms of vortex-line tilting. When the fluid exits the pipe,
the vorticity vector is everywhere azimuthal, and the vortex
lines are concentric circles around the center of the pipe. As the fluid
moves away from the pipe exit, the vortex lines
are gradually tilted by the action of gravity. This imparts to the vorticity a small
horizontal component, which must be compensated to ensure that 
the total horizontal component remains zero. The compensation
occurs by means of an induced recirculation comprising two
counterrotating Dean vortices on either side of the central
symmetry plane of the jet.  

To quantify the foregoing argument, we recall that the axial velocity $u_x(r)$ of the fluid
exiting the pipe is given by (\ref{poisic}). The vorticity associated
with this velocity profile is $(4 U r/a^2)\mathbf e_{\theta}\equiv\omega(r) \mathbf e_{\theta}$, where
$\mathbf e_{\theta}$ is an azimuthal unit vector. Therefore $\omega\sim U/a$ because $r\sim a$.
Now as the fluid moves away from the pipe exit, gravity tilts the vortex lines by
a small angle $\theta (x)$. This gives rise to a small axial component of vorticity
$\sim \pm (U/a)\theta$, where $\pm$ indicates that the axial component has opposite
signs over the right and left halves of the jet section. This vorticity must be compensated
by an axial component of vorticity $\sim \mp U_{Dean}/a$ associated with Dean
recirculation, where $U_{Dean}$ is the
amplitude of that recirculation. Thus we have $U_{Dean}/U\approx \theta$.
To estimate $\theta$, we consider the ballistic trajectory of a particle with
horizontal velocity $\approx U$, which is $y = g x^2/(2 U^2)$. The slope
of that trajectory is $\mathrm dy/\mathrm d x = g x/U^2\approx \theta$. Now
define $x = \xi a$, where $\xi$ is the dimensionless distance from the pipe
exit. Recalling the definitions of $Re$ and $De$, we find 
$U_{Dean}/U \approx \xi (De/Re)^2$. For the case shown in fig. \ref{fig_recirculation},
$Re = 245$, $De = 58$ and $\xi = 2$, which gives $U_{Dean}/U\approx 0.11$.
This value is comparable to the total amplitude $U_{Dean}/U\approx 0.16$
found in the numerical simulation of fig. \ref{fig_recirculation}.

\section{Discussion}

Returning to Torricelli's discovery of the parabolic trajectory of
water jets, we can now see why the simple behavior he observed
is so different from the one we have investigated here. The fundamental difference 
between the two situations is that the velocity profile across a
(turbulent) water jet issuing from a hole is nearly constant, 
whereas the profile across a (laminar) jet issuing from a long pipe is parabolic. 
In consequence, all particles in a water jet follow nearly the same ballistic
(parabolic) trajectory, whereas particles issuing from a laminar flow in a pipe follow different
trajectories depending on their different initial velocities. 
Torricelli's curtain is therefore only observed if the flow through
the pipe is laminar; as soon as it becomes turbulent, we recover the 
situation of a water jet issuing from a hole. The dramatic change in jet
morphology that occurs at the laminar/turbulent transition is clearly visible in the 
NCFMF film ``Turbulence" cited in the Introduction.

Our corrected ballistic model assumes that the velocity
of a particle is the composition of the (dominantly horizontal) motion in an axisymmetric jet
and a vertical motion due to free fall under gravity. This assumption is valid if\
the typical horizontal velocity $\approx U$ of the axisymmetric jet is
greater than or comparable to the free-fall velocity $V$ everywhere within the
`developing' portion of the axisymmetric jet, i.e. before the plug-flow velocity
profile has been achieved. The latter occurs at a distance $x \approx 0.25 a Re\equiv x_0$ from
the pipe exit. The characteristic free-fall velocity at this distance is $V\approx g x_0/U$. 
Using the definitions of $De$ and $Re$, we find that
\be
\frac{V}{U}\leq 1\quad\mathrm{if}\quad \frac{De^2}{Re}\leq 4.
\label{vucriterion}
\ee
Taking the experiment of fig. \ref{fig_trajectories} as an example, we find 
$De^2/Re = 9.2$, or a factor 2.3 larger than the criterion (\ref{vucriterion}).
Thus we conclude that the experiment of fig. \ref{fig_trajectories} 
does not quite satisfy our velocity composition assumption. By contrast, the 
assumption is strongly violated for the experiments with $Re = 120$
and 180 in fig. \ref{fig_shape_vs_re}.

A question raised by our work is whether the distinct secondary 
jet we observe is a consequence of surface tension. Our experiments
cannot by themselves answer this question because
surface tension cannot be eliminated in our setups. By contrast, the
effect of surface tension is negligible in
experiments on miscible buoyant jets injected horizontally into a fluid of comparable 
viscosity \citep{arakeri00jfm, shao17ijhmt}. Because a secondary jet is clearly
seen in a number of these experiments, we conclude that surface tension is not
a necessary condition for its existence. However, if surface tension
is present it will of course influence the position of the secondary jet by 
causing Taylor-Culick retraction of the jet's free edge.

Figure \ref{fig_silicone37_gerris_hxy} 
has shown that our direct numerical simulations of Torricelli's curtain disagree with experiment
in at least one important respect: the vertical extent of the curtain is significantly
smaller in the simulations than in reality. A possible reason for this is limited 
numerical resolution. In our most highly resolved simulations, the smallest
cube in the grid has dimension $a/64$, or 0.13 mm for 
Setup 2 with $a = 8.5$ mm.  For
typical curtain thicknesses $h= 0.5$-1.0 mm measured using this setup
(fig. \ref{fig_hprofiles}),  there are therefore 2 to 4 grid cubes across
the half-thickness $h/2$, which should be adequate to resolve
the structure of the curtain. To verify this, we 
compared a simulation with resolution $a/64$ to one with lower resolution
$a/32$, and found that the vertical extent of the curtain was the same 
to within a few percent. 

A second possible reason for the discrepancy is that our simulations were not 
run long enough to reach steady state. However, this 
is unlikely in view of the test we reported previously in which we
found that simulations run for different times gave indistinguishable results. 

A final possible reason for the disagreement  between simulation and
experiment
is a difference in the effective upstream boundary condition at the 
pipe exit. In the simulations, this condition is a radially symmetric 
parabolic velocity profile corresponding to developed Poiseuille flow
in the pipe. In the laboratory, however, this simple condition may not
be realized. Evidence for this is the non-horizontality of the jet
just after its exit from the pipe, clearly visible in fig. \ref{fig_structure}.
This feature may be due to retroaction of the external
low-pressure ambient air on the fluid in the pipe just upstream
of the exit, modifying the parabolic profile and imparting
a downward component to the exit velocity. Such a downward 
component would have a stronger influence on the relatively 
slow-moving secondary jet than on the primary jet,  which 
is consistent with fig. \ref{fig_silicone37_gerris_hxy}. 
In summary, we believe
that different effective boundary conditions are the most likely
explanation for the discrepancy between experiment
and simulation evident in fig. \ref{fig_silicone37_gerris_hxy}.

Torricelli's curtain has remarkable similarities with the flow in
curved rigid pipes\citep{dean27philmag, berger83annrev}. To investigate flow 
in a pipe whose axis has a 
radius of curvature $R$, one first seeks a scaling for the axial arclength
coordinate $s$ and the transverse (in-section) velocity $\mathbf u_{trans}$
that renders the centrifugal force terms in the transverse momentum
equations comparable in magnitude to the inertial and viscous terms. The
scaling that does this is $s\sim a\delta^{-1/2}$, $\mathbf u_{trans}\sim \delta^{1/2} U$
where $\delta = a/R$. Now for a free horizontal jet whose curvature is due to gravity,
the ballistic equation $y = g x^2/(2 U^2)$ implies $R = U^2/g$ at $x=0$. 
The axial length scale $a\delta^{-1/2}$ then takes the form
$U (a/g)^{1/2}$, which is just the expression for the Dean length $L_D$
that we introduced in \S~\ref{sec_dimanal}. Furthermore, the Dean 
number for a curved pipe is defined as $De = \delta^{1/2} Re$, 
which for our problem turns out to be $De = (a^3 g/\nu^2)^{1/2}$. The
expression $De = \delta^{1/2} Re$ shows that the Dean number is
an effective or `reduced' Reynolds number for flow in a curved pipe.
In fact, in the so-called `loosely coiled pipe' limit $\delta\ll 1$,  $De$ 
is the only 
dimensionless parameter that appears in the rescaled Navier-Stokes
equations, implying that all curved pipe flows with the same value of
$De$ are dynamically similar\citep{dean28philmag}. 
For our experiments using Setup 1, 
$\delta\equiv a g/U^2\in [0.020, 0.52]$, which spans the range
from the loosely coiled (e.g. fig. \ref{fig_shape_vs_re}e with $\delta = 0.02$)
to the strongly coiled limits (e.g. fig. \ref{fig_shape_vs_re}a with $\delta = 0.20$). 
Of course a liquid curtain differs from classical curved pipe flow in
important respects: the outer surface of the fluid is free and
deformable, and surface tension plays an important role in the dynamics.  
Nevertheless, it is illuminating to regard Torricelli's curtain as 
a free-surface curved pipe flow problem in which the shape of the `pipe' 
must be determined as part of the solution. 

In closing, we return to the lava firehose observed at Kilauea volcano, Hawaii in January and February 2017 
(Fig. \ref{fig_hawaii}). Judging by the color of the (basaltic) lava, its temperature $\approx 1200^{\circ}$ C,
whence its kinematic viscosity $\nu\approx 0.012$ m$^2$\,s$^{-1}$ and its density $\rho\approx 2500$ kg\,m$^{-3}$
\citep{villeneuve08chemgeol}. The coefficient of surface tension is 
$\gamma = 0.37$ N\,m$^{-1}$ \citep{walker81contminpet}. From fig. \ref{fig_hawaii} we estimate $a \approx 1$ m,
and assuming a ballistic trajectory for the upper jet we find $U\approx 3$ m s$^{-1}$. To obtain
an independent upper bound on $U$, we used videos to estimate the velocity $V_{ad}$ at which holes
in the curtain are advected downstream, and found $V_{ad} \approx 7$ m s$^{-1}$.  Not surprisingly,
$V_{ad} > U$ because $V_{ad}$ includes a component of vertical motion due to free fall. 
The foregoing values of $U$, $a$, $\nu$ and $\gamma$ imply 
$Re\approx 250$,  $De\approx 260$ and $La\approx 1$. While $La\approx 1$ is much smaller
than in our experiments, the values of the more important parameters $Re$ 
and $De$ are of the same order. We therefore conclude that lava firehoses are natural
examples of Torricelli's curtain, an intriguing phenomenon that to our knowledge has not
previously been studied in depth.


%
%

%

\begin{acknowledgments}
We thank R. Pidoux and L. Auffray for their help in the construction
of the experimental Setup 2. D. Neuville and Y. Ricard kindly provided
values of the physical properties of basaltic lavas.
Two anonymous referees and an associate editor
provided constructive comments and suggestions.
N.M.R. acknowledges support from grant BFC 221950 
from the Progamme National de Plan\'etologie (PNP) of the Institut des Sciences de l'Univers 
(INSU) of the CNRS, co-funded by CNES.
\end{acknowledgments}

\section{Data availability statement}

The data that support the findings of this study are available from the corresponding author
upon reasonable request.

\bibliography{ribe}

\end{document}